%% file: tasi.tex
\documentclass{ws-procs9x6}
\usepackage{epsfig}
\input{mydefs.tex}

\def\bi{\begin{itemize}}
\def\ei{\end{itemize}}
\def\eslt{\not\!\!{E_T}}
\def\esl{\not\!\!{E}}

\def\pbar{\bar{p}}
\def\Dbar{\bar{D}}

\newcommand\prd[3]{{\it Phys.\ Rev.\ }{\bf D #1} (#2) #3}
\newcommand\prep[3]{{\it Phys.\ Rept.\ }{\bf #1} (#2) #3}
\newcommand\prl[3]{{\it Phys.\ Rev.\ Lett.\ }{\bf #1} (#2) #3}
\newcommand\plb[3]{{\it Phys.\ Lett.\ }{\bf B #1} (#2) #3}
\newcommand\jhep[3]{{\it J. High Energy Phys.\ }{\bf #1} (#2) #3}
\newcommand\app[3]{{\it Astropart.\ Phys.\ }{\bf #1} (#2) #3}

\newcommand\ijmpd[3]{{\it Int.\ J.\ Mod.\ Phys.\ }{\bf D #1} (#2) #3}
\newcommand\npb[3]{{\it Nucl.\ Phys.\ }{\bf B #1} (#2) #3}
\newcommand\epjc[3]{{\it Eur.\ Phys.\ J. }{\bf C #1} (#2) #3}
\newcommand\ptp[3]{{\it Prog.\ Theor.\ Phys.\ }{\bf #1} (#2) #3}
\newcommand\zpc[3]{{\it Z.\ Physik }{\bf C #1} (#2) #3}
\newcommand\cpc[3]{{\it Comput.\ Phys.\ Commun.}{\bf #1} (#2) #3}

\newcommand\arnps[3]{{\it Ann.\ Rev.\ Nucl.\ Part.\ Sci.}{\bf  #1} (#2) #3}

\newcommand\ppnp[3]{{\it Prog.\ Part.\ Nucl.\ Phys.}{\bf  #1} (#2) #3}
\newcommand{\hepph}[1]{hep-ph/#1}
\newcommand{\astroph}[1]{astro-ph/#1}

\begin{document}

\title{TASI 2008 lectures on Collider Signals II:\\ 
missing $E_T$ signatures and dark matter connection}

\author{Howard Baer$^*$}

\address{Homer L. Dodge Department of Physics and Astronomy, \\ University of Oklahoma,
Norman, OK 73019, USA\\
$^*$E-mail: baer@nhn.ou.edu}

\begin{abstract}
These lectures give an overview of aspects of missing $E_T$ signatures
from new physics at the LHC, 
along with their important connection to dark matter physics.
Mostly, I will concentrate on supersymmetric (SUSY) sources of $\eslt$,
but will also mention Little Higgs models with $T$-parity (LHT) and
universal extra dimensions (UED) models with $KK$-parity.
Lecture 1 covers SUSY basics, model building and spectra computation. 
Lecture 2 addresses
sparticle production and decay mechanisms at hadron colliders and 
event generation. Lecture 3 covers SUSY signatures at LHC, along with LHT and
UED signatures for comparison. 
In Lecture 4, I address the dark matter connection, and how direct and 
indirect dark matter searches, along with LHC collider searches, 
may allow us to both discover and characterize dark matter in the 
next several years. Finally, the interesting scenario of Yukawa-unified
SUSY is examined; this case works best if the dark matter turns out to be a mixture of
axion/axino states, rather than neutralinos.

\end{abstract}


\bodymatter

\section{Introduction: $\eslt$ collider signatures and the dark matter connection}\label{sec:intro}

I have been assigned the topic of discussing missing transverse energy signatures at hadron colliders,
especially the CERN LHC, which should begin producing data in 2009. 
Every collider event-- be they lepton-lepton, lepton-hadron or hadron-hadron interactions-- 
will contain some amount of missing energy ($\esl$) just due to the fact that 
energy measurements are not perfectly precise. Hadronic events will usually contain more $\esl$
than leptonic events since the energy resolution of hadronic calorimeters is not as
precise as is electromagnetic resolution. In addition, collider detectors are imperfect devices,
often containing un-instrumented regions, cracks, hot or dead calorimeter cells, 
and only partial coverage
of the $4\pi$ steradians surrounding the interaction region: 
always, an allowance must be made at least 
to allow the beam pipe to enter the detector. Also, missing energy can arise from 
cosmic rays entering the detector, or beam-gas collisions, multiple scattering etc. 

At $e^+e^-$ colliders, $\esl$ can be directly measured since the incoming beam 
energies are well-known (aside from bremsstrahlung/beamstrahlung effects at very high energy 
$e^+e^-$ colliders).
At $p\bar{p}$ or $pp$ colliders, the hard scattering events which are likely to produce
new physics are initiated by quark-quark, quark-gluon, or gluon-gluon collisions, and the 
partonic constituents of the proton carry only a fraction of the beam energy. The proton 
constituents not participating in the hard scattering will carry an unknown fraction of beam energy
into the un-instrumented or poorly instrumented forward region, so that for hadron colliders,
only missing {\it transverse} energy ($\eslt$) is meaningful. 
In these lectures, we will concentrate on $pp$ colliders, in anticipation of the first 
forthcoming collisions at the CERN LHC.

Along with these imperfect detector effects, $\eslt$ can arise in Standard Model (SM) production
processes wherein neutrinos are produced, either directly or via particle decays. For instance, 
the key signature for $W$ boson production at colliders was the presence of a hard electron or
muon balanced by missing energy coming from escaping neutrinos produced in  $W^-\to\ell^-\bar{\nu}_\ell$ 
(or charge conjugate reaction) decay (here, $\ell =e$, $\mu$ or $\tau$). In addition, 
neutrinos will be produced by semi-leptonic decays of heavy flavors and $\tau$s, and indeed
$\eslt$ was an integral component of $t\bar{t}$ production events followed by
$t\to bW$ decays at the Fermilab Tevatron, and led to discovery of the top quark.

In these lectures, we will focus mainly on {\it new physics} reactions which lead to
events containing large amounts of $\eslt$. The major motivation nowadays that 
$\eslt$ signatures from new physics should appear at LHC comes from cosmology. 
A large array of data-- coming from measurements of galactic rotation curves, 
galaxy cluster velocity profiles, gravitational lensing, hot gas in galaxy clusters,
light element abundances in light of Big Bang nucleosynthesis, large scale structure
simulations and measurements
and especially measurements of anisotropies in the cosmic microwave background radiation (CMB)--
all point to a consistent picture of a universe which is constructed of
\begin{itemize}
\item baryons: $\sim 4\%$
\item dark matter: $\sim 21\%$
\item dark energy: $\sim 75\%$
\item $\nu$s, $\gamma$s: a tiny fraction.
\end{itemize}
Thus, particles present in the SM of particle physics constitute only $\sim 4\%$ of the 
universe's energy budget, while the main portion is comprised of dark energy 
(DE) which causes an accelerating expansion of the universe and dark matter (DM).
There is an ongoing program of experiments under way or under development which will
probe the dark energy. The goal is to try to tell if DE comes from Einstein's cosmological
constant (which theoretical prejudice says really ought to be there), or something quite different.
Measurements focus on the parameters entering the dark energy equation of state. 

For dark matter, there exist good reasons to believe that it will be connected to
new physics arising at the weak scale: exactly the energy regime to be probed by the LHC.
The density of dark matter is becoming known with increasing precision.
The latest WMAP5 measurements\cite{wmap5} find
\be
\Omega_{CDM}h^2=0.110\pm 0.006
\label{eq:wmap}
\ee
where $\Omega_{CDM}$ is the cold DM density divided by the critical closure density of the universe
$\rho_c$, and $h$ is the scaled Hubble constant: $h=0.73^{+0.04}_{-0.03}$.
While the cosmic DM density is well-known, the identity
of the DM particle(s) is completely unknown. Nevertheless, we do know some properties of the
dark matter: it must have mass, it must be electrically neutral (non-interacting with light)
and likely color neutral, and it must have a non-relativistic velocity profile, 
{\it i.e.} be what is termed {\it cold} dark matter (CDM). The reason it must be cold
is that it must seed structure formation in the universe, {\it i.e.} it must clump.
That means its velocity must be able to drop below its escape velocity; otherwise, it would 
disperse, and not seed structure formation.
We already know of one form of dark matter: cosmic neutrinos. However, the SM neutrinos
are so light that they must constitute {\it hot} dark matter.

\begin{figure}[tbh]
\begin{center}
\includegraphics[width=6.5cm]{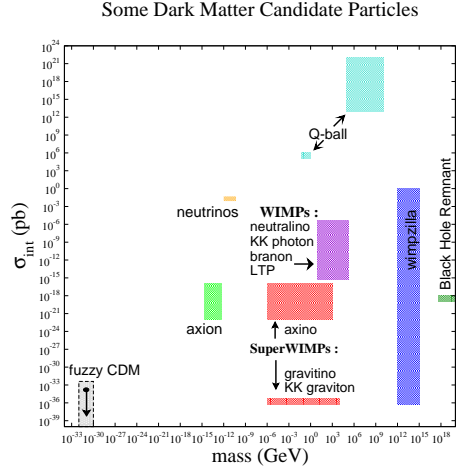}
\end{center}
\vspace{-8mm}
\caption{\small\it 
Dark matter candidates in the mass versus interaction 
strength plane, taken from Ref.\cite{dmsag}:
http://www.science.doe.gov/hep/hepap\_reports.shtm. 
See also, L. Roszkowski, Ref.\cite{cdm_review}. }
\label{fig:DM} 
\end{figure}

The theoretical literature contains many possible candidate DM particles. A plot of some of these 
is shown in Fig. \ref{fig:DM} (adapted from L. Roszkowski) 
in the mass vs. interaction cross section plane.
Some of these candidates emerge from attempts to solve longstanding problems in particle
physics. For instance, the axion arises from the Peccei-Quinn solution to the strong $CP$
problem. The weakly interacting massive particles (WIMPs) frequently arise from models
which attempt to explain the origin of electroweak symmetry breaking. WIMP dark matter
candidates include the lightest neutralino of models with weak scale supersymmetry,
while Kaluza-Klein photons arise in models with universal extra dimensions, and 
lightest $T$-odd particles arise in Little Higgs models with a conserved $T$-parity.

\vspace{-6mm}

WIMP particles have an additional motivation, which has recently been coined the
``WIMP miracle''\cite{fk}. The idea here is to assume the existence of a dark matter particle
which was once in thermal equilibrium at high temperatures in the early universe.
The equilibrium abundance is easily calculated from thermodynamics and provides one boundary condition.
As the universe expands and cools, the DM particle will drop out of thermal
equilbrium (freeze-out); while DM will no longer be produced thermally, it can still annihilate
away. The subsequent time evolution of the dark matter density is governed by the Boltzmann
equation as formulated for the Friedman-Robertson-Walker universe:
\be
dn/dt = -3Hn-\langle \sigma v_{rel}\rangle (n^2-n_0^2)
\ee
wherein the $-3Hn$ terms gives rise to dilution due to the Hubble expansion,
and the $\langle \sigma v_{rel}\rangle$ term involves the dark matter
annihilation cross section.
The Boltzmann equation is solved in various textbooks\cite{kolb_turner},
and the solution is given approximately by
\be
\Omega h^2=\frac{s_0}{\rho_c/h^2}\left(\frac{45}{\pi g_*}\right)^{1/2}\frac{x_f}{M_{Pl}}
\frac{1}{\langle \sigma v\rangle} ,
\ee
where $s_0$ is the current entropy density, $g_*$ is the effective degrees of freedom and 
$x_f$ is the freeze-out temperature $\sim m_{\tz_1}/20$.
The quantities entering ths expression are unrelated to one another and several have large
exponents. Filling in numerical values, they conspire to give 
\be
\Omega h^2\sim \frac{0.1\ {\rm pb}}{\langle\sigma v\rangle},
\ee
so that in order to obtain the measured dark matter abundance $\Omega h^2\sim 0.11$, 
we expect $\langle \sigma v\rangle $ to be a {\it weak scale sized cross section}!.
Inputing a typical weak scale cross section, 
\be
\sigma =\frac{\alpha^2}{8\pi}\frac{1}{m^2}
\ee
 and setting $\sigma\sim 1$ pb, we find $m\sim 100$ GeV!  
The lesson here is an important one: that the measured dark matter density is
consistent with the annihilation cross section expected of a weakly interacting particle
with {\it mass of order the weak scale}. This auspicious relation amongst 
dark matter density, annihilation cross section and particle mass is sometimes refered
to as the {\it WIMP miracle}, and seemingly offers independent {\it astrophysical}
evidence for new physics at or around the weak scale: in this case the new
physics is associated with the cosmic density of dark matter. 
If such a WIMP particle exists, then it should be produced at large rates
at the CERN LHC. Since it is dark matter, once produced in a collider experiment,
it would escape normal detection. If produced in association with other SM 
particles (such as quarks and gluons), then it will carry away 
missing (transverse) energy, and will give rise to new physics signatures including  
jets plus $\eslt$.
It is this connection of weak scale physics and dark matter which makes the 
jets $+\eslt$ signature so important and exciting at the LHC.

\section{Lecture 1: SUSY basics: model building and spectra calculation}
\label{sec:lec1}

\subsection{The Wess-Zumino model}

We begin with a toy model due to Wess and Zumino, from 1974\cite{wz}.
The model Lagrangian is given by
\be
{\cal L}={\cal L}_{\rm kin.}+{\cal L}_{\rm mass}
\ee
with
\bea
{\cal L}_{\rm kin.}&=&{1\over 2}(\partial_{\mu}A)^2+{1\over 2}(\partial_{
mu}B)^2+ {i\over 2}\overline{\psi}\dsl\psi+{1\over 2}(F^2+G^2)\\
{\cal L}_{\rm mass}&=&-m[ {1\over 2}\psib\psi -GA-FB ],
\eea
where $A$ and $B$ are real scalar fields, $\psi$ is a 4-component
Majorana spinor field and $F$ and $G$ are non-propagating {\it auxiliary}
fields. The auxiliary fields may be eliminated from the Lagrangian via the
Euler-Lagrange equations: $F=-mB$ and $G=-mA$.

But instead, impose the transformations
\bea 
A&\to  &A+\delta A\ {\rm with}\ \delta A =i\bar{\alpha}\g5\psi ,\\
\delta B &=&-\bar{\alpha}\psi,\\
\delta\psi &=&-F\alpha+iG\g5\alpha+\dsl\g5 A\alpha +i\dsl B\alpha ,\\
\delta F &=&i\bar{\alpha}\dsl \psi,\\
\delta G &=&\bar{\alpha}\g5 \dsl \psi .
\eea
Note that the transformation mixes bosonic and fermionic fields. It is 
called a {\it supersymmetry} transformation.
Using Majorana bilinear re-arrangements 
(such as $\psib\chi=-\bar{\chi}\psi$), the product rule
$\dmu (f\cdot g) =\dmu f\cdot g+f\cdot \dmu g$ and some algebra, we find that
$\CL\to\CL +\delta\CL$ with
\bea
\delta\CL_{\rm kin} &=&                                                   
\partial^{\mu}(-\frac{1}{2}\ab\gamma_{\mu}\dsl B\psi                            
+\frac{i}{2}\ab\g5\gmu\dsl A\psi                                                
+\frac{i}{2}F\ab\gmu\psi +\frac{1}{2}G\ab\g5\gmu\psi),\\
\delta\CL_{\rm mass}& =&\partial^\mu (mA\ab\g5\gmu\psi +imB\ab\gmu\psi ) .
\eea
Since the Lagrangian changes by a total derivative, the {\it action}
$S=\int {\cal L}d^4x$ is invariant! (owing to Gauss' law in 4-d
$\int_V d^4x \dmu \Lambda^\mu =
\int_{\partial V} d\sigma \Lambda^\mu n_\mu$).
Under supersymmetry transformations, 
the Lagrangian always transforms into a total derivative, leaving the 
action invariant.

The Wess-Zumino (WZ) model as presented in 1974 exhibited a qualitatively 
new type of symmetry transformation with astonishing properties: one being that
the usual quantum field theory quadratic divergences associated with scalar fields
all cancel! Theories with this sort of property might allow for a 
solution of the so-called ``gauge hierarchy problem'' of 
grand unified theories (GUTs). 

\subsection{Superfield formalism}

While the WZ model was very compelling, it was constructed by hand,
and offered no insight into how to construct general supersymmetric models.
A short while later, Salam and Strathdee\cite{ss} developed the superfield formalism,
which did provide rules for general SUSY model building. They first 
expanded the usual four-component spacetime $x^\mu$ to {\it superspace}:
$x^\mu \to (x^\mu,\theta_a )$, where $\theta_a$ ($a=1-4$) represents four
{\it anti-commuting } (Grassmann vaued) dimensions arranged as a Majorana spinor.
Then they introduced {\it general} superfields:
\bea
\hat{\Phi}(x,\theta )&=& \CS-i\sqrt{2}\bar{\theta}\g5\psi -{i\over 2}
(\bar{\theta}\g5\theta)\CM +{1\over 2}(\bar{\theta}\theta )
\CN +{1\over 2}(\bar{\theta}\g5\gmu\theta )V^\mu \nonumber \\
&+& i(\bar{\theta}\g5\theta )[\bar{\theta}(\lambda+{i\over \sqrt{2}}\dsl\psi )]
-{1\over 4}(\bar{\theta}\g5\theta )^2[\CD-{1\over 2}\Box\CS ],
\eea
left chiral scalar superfields (LCSSF):
\be
\Shat_L (x,\th )=\CS (\xhat )+i\sqrt{2}\thb\psi_L(\xhat )                      
+i\thb\th_L\CF (\xhat )
\ee 
(where $\xhat_\mu =x_\mu +{i\over 2}\thb\g5\gmu\th$)
and right chiral scalar superfields (RCSSF):
\be
\Shat_R (x,\th )=\CS ({\xhat}^\dagger )-i\sqrt{2}\thb\psi_R                    
({\xhat}^\dagger) -i\thb\th_R\CF ({\xhat}^\dagger) .
\ee
The multiplication table for superfields is as follows:
\bi
\item $LCSSF\times LCSSF= LCSSF$
\item $RCSSF\times RCSSF=RCSSF$
\item $LCSSF\times RCSSF=$ a general superfield.
\ei
Moreover, under a supersymmetry transformation, 
the $D$-term of a general superfield and the $F$ term of a LCSSF or a 
RCSSF always transforms into a total derivative.

This last observation gave the key insight allowing for construction of  general supersymmetric 
models. The superpotential $\hat{f}$ is a function of LCSSFs only, and hence by the
multiplication table above is itself a LCSSF.
Its $F$ term transforms into a total derivative under SUSY, and hence is a
candidate Lagrangian. Also, the Kahler potential $K$ is introduced: it is a function of
products of RCSSFs times LCSSFs: it is therefore a general superfield and 
its $D$-term transforms into a total derivative under SUSY, and hence its $D$-term 
is a candidate Lagrangian. The choice $K=\sum_{fields} \hat{\cal S}^\dagger \hat{\cal S}$
is completely general for a renormalizable theory.
Augmenting the above superfields with {\it gauge} 
superfields (ones containing spin-1 gauge bosons and spin-$1\over 2$ 
gauginos all transforming under the adjoint representation) 
allows one to construct locally gauge invariant, renormalizable
field theories which are also invariant under SUSY transformations. 

A Master Lagrangian can then be constructed. Its form is given in
Ch. 5 of Ref. \cite{wss}. It allows for a recipe for SUSY model building:
\bi
\item select the gauge group (simple or direct product),
\item select the matter and Higgs representations,
\item write down a superpotential (gauge invariant and renormalizable),
\item plug all terms into the Master Lagrangian, calculate the physical
states and Feynman rules, and you are good to go!
\ei 

\subsection{Minimal Supersymmetric Standard Model (MSSM)}
 
The superfield formalism\cite{wss,dgr,bine} allows for the construction of a
supersymmetric version of the Standard Model, known as the Minimal
Supersymmetric Standard Model, or MSSM. For each
quark and lepton of the SM, the MSSM necessarily includes spin-0
superpartners $\tq_L$ and $\tq_R$ along with $\tell_L$ and $\tell_R$,
whose gauge quantum numbers are fixed to be the known gauge quantum
numbers of the corresponding SM fermions. Thus, for example, the
right-handed up quark scalar (usually denoted by $\tu_R$) is a
color-triplet, weak isosinglet with the same weak hypercharge 4/3 as the
right-handed up-quark.  The MSSM thus includes a plethora of new scalar
states: $\te_L$, $\te_R$, $\tnu_{eL}$, $\tu_L$, $\tu_R$, $\td_L$,
$\td_R$ in the first generation, together with analogous states for the
other two generations.  Spin-zero {\it squark} partners of quarks with
large Yukawa couplings undergo left-right mixing: thus, the $\tst_L$ and
$\tst_R$ states mix to form mass eigenstates -- $\tst_1$ and $\tst_2$ --
ordered from lowest to highest mass.  

The spin-0 Higgs bosons are
embedded in Higgs superfields, so that the MSSM also includes spin-$1\over 2$
higgsinos.  Unlike in the SM, the same Higgs doublet cannot give a mass
to both up- and down- type fermions without catastrophically
breaking the underlying supersymmetry.
Thus  the MSSM includes two Higgs doublets instead of one as in the SM.  
This gives rise to a richer
spectrum of physical Higgs particles, including neutral light $h$ and heavy $H$
scalars, a pseudoscalar $A$ and a pair of charged Higgs bosons $H^\pm$.

The gauge sector of the MSSM contains gauge bosons along with spin-half
gauginos in the adjoint representation of the gauge group: thus, along
with eight colored gluons, the MSSM contains eight colored spin-$1\over
2$ gluinos. Upon electroweak symmetry breaking, the four gauginos of
$SU(2)_L\times U(1)_Y$ mix (just as the $SU(2)_L$ and $U(1)_Y$ gauge
bosons mix) amongst themselves {\it and} the higgsinos, to form
charginos -- $\tw_1^\pm$ and $\tw_2^\pm$ -- and  neutralinos --
$\tz_1$, $\tz_2$, $\tz_3$ and $\tz_4$.  The $\tz_1$ state, the lightest
neutralino, is often the lightest supersymmetric particle
(LSP), and turns out to be an excellent WIMP candidate for CDM in the
universe.

If nature is perfectly supersymmetric, then the {\it spin-{\rm 0}
superpartners would have exactly the same mass} as the corresponding
fermions. Charged spin-0 partners of the electron with a mass of
0.51~MeV could not have evaded experimental detection. Their non-observation
leads us to conclude that SUSY must be a {\it broken}
symmetry. In the MSSM, SUSY is broken {\it explicitly} by including so-called
soft SUSY breaking (SSB) terms in the Lagrangian. The SSB terms preserve
the desirable features of SUSY, such as the stabilization of the scalar
sector in the presence of radiative corrections, while lifting the
superpartner masses in accord with what is necessary from experiment.
It is important to note that the equality of dimensionless couplings
between particles and their superpartners is still preserved (modulo
small effects of radiative corrections): in particular,
phenomenologically important  gauge
interactions of superpartners and the corresponding interactions of
gauginos remain (largely) unaffected by the SSB terms. 

\subsection{Supergravity}

The addition of the SSB Lagrangian terms may seem {\it ad-hoc} and ugly.
It would be elegant if instead supersymmetry could be spontaneously
broken. But it was recognized in the early to mid-1980's that models
where global SUSY is spontaneously broken at the weak scale ran into
serious difficulties. The situation is very different if we elevate SUSY
from a global symmetry to a {\it local} one. In local SUSY, we are
{\it forced to include the graviton/gravitino super-multiplet} into the
theory, in much the same way that we have to include spin-1 gauge fields
to maintain local gauge invariance of Yang-Mills theories.
Theories with local SUSY are known as {\it supergravity} (SUGRA)
theories because they are supersymmetric and necessarily include
gravity.  Moreover, the gravitational sector of the theory reduces to
general relativity in the classical limit.  Within the framework of
SUGRA, it is possible to add an additional sector whose dynamics
spontaneously breaks SUSY but which interacts with SM particles and
their superpartners only via gravity (the so-called hidden sector). The
spontaneous breakdown of supersymmetry results in a mass for the
gravitino in the same way that in local gauge theories gauge bosons
acquire mass by the Higgs mechanism. This is, therefore, referred to as
the {\it super-Higgs mechanism.} The remarkable thing is that because of
the gravitational couplng between the hidden and the MSSM sectors, 
the effects of spontaneous supersymmetry
breaking in the hidden sector are conveyed to the MSSM sector, and 
(provided the SUSY-breaking scale in the hidden sector is appropriately chosen)
weak scale SSB
terms that lift the undesirable degeneracies between the masses of SM
particles and their superpartners are automatically induced. 
Indeed, in the limit where
$M_{\rm Pl}\to\infty$ (keeping the gravitino mass fixed), we recover a
global SUSY theory along with the desired SSB terms! The gravitino typically
has a weak scale mass and in many cases decouples from collider experiments
because of its tiny gravitational couplings.
For reasons
that we cannot discuss here, these locally supersymmetric models are
free\cite{wss,dgr,bine} of many of the difficulties that 
plague globally supersymmetric models.

\subsection{SUGRA GUTs}

Motivated by the successful unification of gauge couplings at a scale
$M_{\rm GUT}\sim 2\times 10^{16}$ GeV in the MSSM, we are led to
construct a grand unified theory (GUT) based on local supersymmetry. 
In this case, the theory
renormalized at $Q=M_{\rm GUT}$ contains just one gaugino mass parameter
$m_{1/2}$. Renormalization effects then split the physical gaugino
masses in the same way the measured values of the gauge couplings
arise from a single unified GUT scale gauge coupling. In general,
supergravity models give rise to complicated mass matrices for the
scalar superpartners of quarks and leptons, with concomitant flavor
violation beyond acceptable levels. However, in models with 
{\it universal} soft SUSY breaking terms, a super-GIM mechanism suppresses
flavor violating processes\cite{dimop}.  In what has come to be known as
the minimal supergravity (mSUGRA) model, a universal scalar mass $m_0$
and also a universal SSB scalar coupling $A_0$ are assumed to exist at a
high scale $Q\sim  M_{\rm GUT}-M_{\rm Pl}$.  The physical masses of squarks
and sleptons are split after renormalization, and can be calculated
using renormalization group techniques. Typically, in the mSUGRA model,
we have $m_{\tq} \agt m_{\tell_L} \agt m_{\tell_R}$.  Although the Higgs
scalar mass parameters also start off at the common value $m_0$ at the
high scale, the large value of the top quark Yukawa coupling drives the
corresponding squared mass parameter to negative values and EWSB is
{\it radiatively broken}.  Within this framework,
the masses and couplings required for phenomenology are fixed by just a
handful of parameters which are usually taken to be, 
\be 
\ \ \ m_0,\ m_{1/2},\ A_0,\ \tan\beta ,\ {\rm and}\ sign(\mu ) .  
\ee 
Here,
$\tan\beta$ is the ratio of the vacuum expectation values of the Higgs
fields that give masses to up and down type fermions, and $\mu$ is the
supersymmetric higgsino mass parameter whose magnitude is fixed to
reproduce the measured value of $M_Z$.  If all parameters are real, then
potentially large $CP$-violating effects are suppressed as well.
Computer codes such as Isajet, SuSpect, SoftSUSY and Spheno that
calculate the full spectrum of sparticle and Higgs boson masses are
publicly available\cite{kraml}.

\subsection{Some realistic SUSY models}

The mSUGRA model (sometimes referred to as the constrained MSSM or
CMSSM) serves as a paradigm for many SUSY phenomenological
analyses. However, it is important to remember that it is based on many
assumptions that can be tested in future collider experiments but which
may prove to be incorrect.  For instance, in many GUT theories, it is
common to get {\it non-universal} SSB parameters.  In addition, there
are other messenger mechanisms besides gravity.  In gauge-mediated SUSY
breaking models (GMSB)\cite{gmsb}, a special messenger sector is
included, so gravitinos may be much lighter than all other sparticles,
with implications for both collider physics and cosmology. In
anomaly-mediated SUSY breaking (AMSB) models\cite{amsb}, gravitational
anomalies induce SSB terms, and the gravitino can be much heavier than
the weak scale. There are yet other models\cite{kklt} where SSB
parameters get comparable contributions from gravity-mediated as well
as from anomaly-mediated sources, and very recently, also from
gauge-mediation\cite{EKZ}. The pattern of superpartner masses is
sensitive to the mediation-mechanism, so that we can expect collider
experiments to reveal which of the various mechanisms that have been
proposed are actually realized in nature.  We also mention that in both
the GMSB and AMSB models, it is somewhat less natural (but still
possible!) to obtain the required amount of SUSY dark matter in the
Universe. Although these are all viable scenarios, they have not been as
well scrutinized as the mSUGRA model.

\section{Lecture 2: Sparticle production, decay and event generation}
\label{sec:lec2}

\subsection{The role of the CERN Large Hadron Collider}

The CERN Large Hadron Collider (LHC) turned on briefly in 2008 before
an electrical mis-connection caused a helium leak, and a shut-down until
mid-2009. The LHC is a $pp$ collider, expected to operate ultimately at
center-of-mass energy $\sqrt{s}=14$ TeV. Since protons are not fundamental
particles, we may think instead of LHC as being a quark-gluon collider, where 
the quark and gluon constituents of the proton that undergo hard
strong or electroweak scatterings only carry typically a small
fraction of the proton energy. 

For the hadronic reaction 
\be
A+B\to c+d+X ,
\ee
where $A$ and $B$ are hadrons, $c$ and $d$ might be superpartners, and 
$X$ is associated hadronic debris, we are really interested in the
subproces reaction 
\be
a+b\to c+d ,
\ee
where $a$ is  a partonic constituent of hadron $A$, and $b$ is a partonic constituent
of hadron $B$.  The SM or MSSM or whatever effective theory we are working in 
provides us with the Feynman rules to calculate the subprocess cross-section
$\hat{\sigma} (ab\to cd)$. To find the total hadron-hadron cross section, 
we must convolute the subprocess cross section with the 
{\it parton distribution functions} $f_{a/A}(x_a,Q^2)$ which gives the probability to 
find parton $a$ in hadron $A$ with momentum fraction $x_a$ at energy-squared scale $Q^2$.
Thus, we have
\be
d\sigma (AB\to cd X) =\sum_{a,b}
\int_0^1 dx_a \int_0^1 dx_b f_{a/A}(x_a,Q^2)\ f_{b/B}(x_b,Q^2)
d\hat{\sigma}(ab\to cd) .
\label{eq:partonmodelcs}
\ee
where the sum extends over all initial partons $a$ and $b$ whose
collisions produce the final state particles $c$ and $d$.
The parton-parton center-of-mass energy-squared $\hat{s}=x_a x_b s$ which enters the hard
scatterng is only a small fraction of the proton-proton center-of-mass energy-squared $s$. 
Thus, to explore the TeV scale, a hadron-hadron collider of tens of TeV is indeed needed!

Around the 17 mile circumference LHC ring are situated four experiments:
Atlas, CMS, LHC-B and Alice. Atlas and CMS are all-purpose detectors
designed to detect almost all the energy emitted in hard scattering reactions.
They will play a central role in the search for new physics at the LHC, and the search
for dark matter production in the LHC environment.

\subsection{Cross section calculations for the LHC}

As noted above, if a perturbative model for new physics is available, then
the relativistic-quantum mechanical amplitude ${\cal M}$ for the relevant
scattering sub-process may be calculated, usually at a low order in 
perturbation theory. The probability for scattering is given 
by the square of the (complex) scattering amplitude ${\cal M}{\cal M}^\dagger$, 
and can often be expressed 
as a function of dot products of the external leg momenta entering the
diagrams, after summing and averaging over spins and colors. Nowadays,
several computer programs-- such as CalcHEP, CompHEP and MadGraph-- are 
available to automatically do tree level calculations of various $2\to n$
processes.

To gain the parton level scattering cross section, one must multiply by suitable phase space
factors and divide by the flux factor. The phase space integrals for 
simple sub-processes may be done analytically. More often, the final integrals are
done using Monte Carlo (MC) techniques, which efficiently allow integration over multi-body
phase space. The MC technique, wherein random integration variables are generated
computationally, and summed over, actually allows for a {\it simulation} of the scattering
process. This can allow one to impose cuts, or simulate detectors, to gain a
more exact correspondence to the experimental environment.

\subsection{Sparticle production at the LHC}

Direct production of neutralino dark matter at the LHC ($pp\to
\tz_1\tz_1 X$, where $X$ again stands for assorted hadronic debris) is of
little interest since the high $p_T$ final state particles all escape
the detector, and there is little if anything to trigger an event
record. Detectable events come from the production of the heavier
superpartners, which in turn decay via a multi-step cascade which
ends in the stable lightest SUSY particle (LSP).

In many SUSY models, the strongly interacting squarks and/or gluinos are
among the heaviest states. Unless these are extremely heavy, they will
have large production cross sections at the LHC. Strong interaction
production mechanisms for their production include, 1. gluino pair
production $\tg\tg$, 2. squark pair production $\tq\tq$ and
3. squark-gluino associated production $\tq\tg$. Note here that the
reactions involving squarks include a huge number of subprocess
reactions to cover the many flavors, types (left- and right-), and also
the anti-squarks. The various possibilities each have different angular
dependence in the production cross sections\cite{bt3}, and the different
flavors/types of squarks each have different decay modes\cite{cascade}. 
These all have to
be kept track of in order to obtain a reliable picture of the
implications of SUSY in the LHC detector environment. 
Squarks and gluinos can also be
produced in association with charginos and neutralinos\cite{assprod}. 
Associated gluino production occurs via squark exchange in the 
$t$ or $u$ channels and is suppressed if squarks are very heavy.

If colored sparticles are very heavy, then electroweak production of charginos
and neutralinos may be the dominant sparticle production mechanism at the
LHC. The most important processes are pair production of charginos,
$\tw_i^\pm\tw_j^\mp$ where $i,j=1,2$, and chargino-neutralino
production, $\tw_i^\pm \tz_j$, with $i=1,2$ and $j=1-4$. In models with
unified GUT scale gaugino masses and large $|\mu|$, $Z\tw_1\tw_1$ and
$W\tz_2\tw_1$ couplings are large so that $\tw_1\tw_1$ and $\tw_1\tz_2$
production occurs at significant rates.  The latter process can lead to
the gold-plated trilepton signature at the LHC\cite{trilhc}. Neutralino
pair production ($pp\to \tz_i\tz_j X$ where $i,j=1-4$) is also possible.
This reaction occurs at low rates at the LHC 
unless $|\mu| \simeq M_{1,2}$ (as in the case of mixed higgsino dark matter (MHDM)). 
Finally, we mention slepton pair production:
$\tell^+\tell^-$, $\tnu_\ell\tell$ and $\tnu_\ell\bar{\tnu}_\ell$, which
can give detectable dilepton signals if $m_{\tell}\alt 300$
GeV\cite{slep}.

\begin{figure}[tbh]
\begin{center}
\includegraphics[width=7cm]{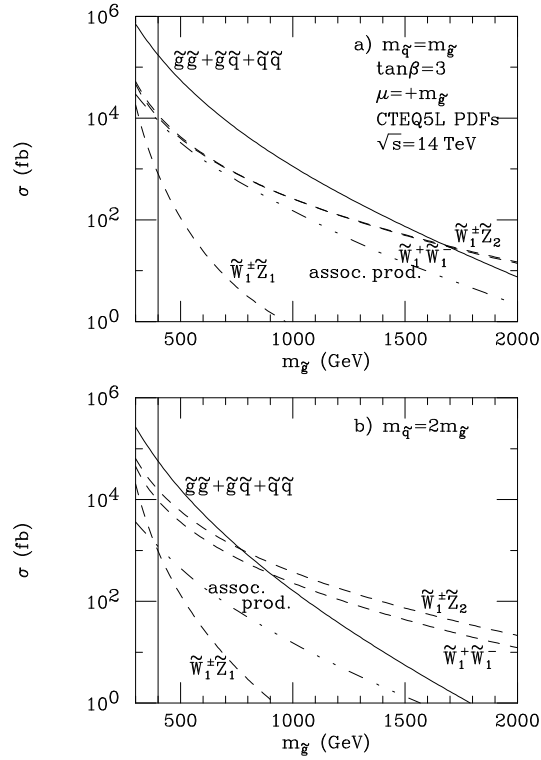}
\end{center}
\vspace{-3mm}
\caption{\small\it 
Cross sections for production of various sparticles
at the LHC. Gaugino mass unification is assumed. }
\label{fig:sigma} 
\end{figure}

In Fig. \ref{fig:sigma} we show various sparticle production cross sections
at the LHC as a function of $m_{\tg}$. Strong interaction
production mechanisms dominate at low mass, while electroweak processes
dominate at high mass. The associated production mechanisms are never
dominant. 
 The expected LHC integrated luminosity in the first year
of running is expected to be around 0.1 fb$^{-1}$, 
while several tens of fb$^{-1}$ of
data is expected to be recorded in the first several years of
operation. The ultimate goal is to accumulate
around 500-1000~fb$^{-1}$, correponding to $10^5-10^6$ SUSY events for
$m_{\tg}\sim$~1~TeV.

\subsection{Sparticle cascade decays}

In $R$-parity conserving SUSY models, sparticles decay to lighter sparticles
until the decay terminates in the LSP\cite{cascade}. 
Frequently, the direct decay to
the LSP is either forbidden or occurs with only a small branching
fraction.  Since gravitational interactions are negligible, gluinos can
only decay via $\tg\to q\tq$, where the $q$ and $\tq$ can be of any
flavor or type. If two body decay modes are closed, the squark will be
virtual, and the gluino will decay via three body modes $\tg\to
q\bar{q}\tz_i,\ q\bar{q}'\tw_j$. If squarks are degenerate, and Yukawa
coupling effects negligible, three-body decays to the wino-like chargino
and neutralino usually have larger branching fractions on account of
the larger gauge coupling. If $|\mu| < M_2$, gluinos and squarks may
thus decay most of the time to the heavier charginos and neutralinos,
resulting in lengthy cascade decay chains at the LHC.

Squarks decay always to two-body modes: $\tq\to q\tg$ if it is
kinematically allowed, or $\tq_L\to q'\tw_i,\ q\tz_j$, while $\tq_R\to
q\tz_j$ only, since right-squarks do not couple to charginos.  Sleptons
do not have strong interactions so cannot decay to gluinos. Their
electroweak decays are similar to corresponding decays of squarks
$\tell_L\to \ell'\tw_i,\ \ell\tz_j$ while $\tell_R\to \ell\tz_j$ only.

Charginos may decay via two-body modes: $\tw_i\to W\tz_j,\
\tell\nu_\ell,\ \ell\tnu_\ell,\ Z\tw_j$ or even to $\phi\tw_j$ or
$H^-\tz_j$, where $\phi =h,H,A$. If two-body modes are inaccessible,
then three-body decays dominate: $\tw_i\to \tz_j f\bar{f}'$, where $f$
and $f'$ are SM fermions which couple to the $W$. Frequently, the decay
amplitude is dominated by the virtual $W$ so that the three-body decays
of $\tw_1$ have the same branching fractions as those of the $W$.
Neutralinos decay via $\tz_i\to W\tw_j,\ H^+\tw_j,\ Z\tz_j,\ \phi\tz_j$
or $f\tf$.  If two body neutralino decays are closed, then
$\tz_i\to\tz_j f\bar{f}$, where $f$ are the SM fermions. In some models, the
branching fraction for radiative decays $\tz_i\to \tz_j\gamma$ (that
only occurs at the one-loop level) may be significant\cite{z2z1g}. 
The cascade decay modes
of neutralinos depend sensitively on model parameters\cite{bt93}.

If $\tan\beta$ is large, then $b$ and $\tau$ Yukawa coupling effects
become important, enhancing three body decays of $\tg$, $\tw_i$ and
$\tz_j$ to third generation fermions\cite{bcdpt}. For very large values
of $\tan\beta$ these decays can even dominate, resulting in large rates
for $b$-jet and $\tau$-jet production in SUSY events\cite{ltanb_lhc}.

Finally, the various Higgs bosons can be produced both directly and via
sparticle cascades at the LHC\cite{hatlhc}. 
Indeed, it may be possible that $h$ is first
discovered in SUSY events because in a sample of events enriched for
SUSY, it is possible to identify $h$ via its dominant $h\to b\bar{b}$
decays rather than via its sub-dominant decay modes, as required for 
conventional searches\cite{hatlhc}. The heavier Higgs bosons decay to a
variety of SM modes, but also to SUSY particles if these latter decays
are kinematically allowed, leading to novel signatures
such as $H,\ A\to\tz_2\tz_2\to 4\ell+\eslt$\cite{Hto4l}.

The cascade decays terminate in the LSP. In the case of a $\tz_1$ LSP,
the $\tz_1$ is a DM candidate, and leaves its imprint via $\eslt$. In
the case of a weak scale $\tG$ or $\ta$ LSP, then $\tz_1$ will
decay as $\tz_1\to \gamma\tG$ or $\gamma \ta$. 
In these cases, the $\tz_1$ lifetime is long
enough that it decays outside the detector, so one still expects large
$\eslt$ in the collider events. An exception arises for the case of
super-light gravitinos (with masses in the eV to keV range) that are
possible in GMSB models: see (\ref{gravdec}). Then, the decay may take
place inside the detector, possibly with a large vertex
separation. It is also possible that the NLSP is charged and
quasi-stable, in which case 
collider events may include highly ionizing tracks instead
of, or in addition to, $\eslt$.

The decay branching fractions depend on the entire spectrum of SUSY
particle masses and and their mixings. They are pre-programmed in several
codes: Isajet\cite{isajet}, SDECAY\cite{sdecay} and Spheno\cite{spheno}.

\subsection{Event generation for LHC}

Once sparticle production cross sections and decay branching fractions
have been computed, it is useful to embed these into event generator
programs to simulate what SUSY collider events will look like at LHC.
There are several steps involved:
\bi
\item Calculate all sparticle pair production cross sections. Once all
initial and final states are accounted for, this involves over a
thousand individual subprocess reactions. In event generation, a
particular reaction is selected on a probabilistic basis, with a weight
proportional to its differential cross-section.

\item Sparticle decays are selected probabilistically into all the 
  allowed modes in proportion to the corresponding 
branching fractions.

\item Initial and final state quark and gluon radiation are usually
dealt with using the parton shower (PS) algorithm, which allows for
probabilistic parton emission based on approximate collinear QCD
emission matrix elements, but exact kinematics. The PS is also applied at
each step of the cascade decays, which may lead to additional jet production
in SUSY collider events.

\item A hadronization algorithm provides a model for turning various
quarks and gluons into mesons and baryons. Unstable hadrons must be
further decayed.

\item The beam remnants -- proton constituents not taking part in the hard
scattering -- must be showered and hadronized, usually with an
independent algorithm, so that energy deposition in the forward 
detector region may be reliably calculated.  
\ei

At this stage, the output of an event generator program is a listing of
particle types and their associated four-vectors. The resulting event
can then be interfaced with detector simulation programs to model what the
actual events containing DM will look like in the environment of a
collider detector.

Several programs are available, including Isajet\cite{isajet},
Pythia\cite{pythia}, Herwig\cite{herwig} and Sherpa{\cite{sherpa}. 
Other programs such as
Madevent\cite{mad}, CompHEP/CalcHEP\cite{comp} and Whizard\cite{whizard}
can generate various
$2\to n$ processes including SUSY particles. The output of these
programs may then be used as input to Pythia or Herwig for showering and
hadronization. Likewise, parton level Isajet SUSY production followed by
cascade decays can be input to Pythia and Herwig via the Les Houches
Event format\cite{lhe}.

\section{Lecture 3: SUSY, LHT and UED signatures at the LHC}
\label{sec:lec3}

\subsection{Signatures for SUSY particle production}

Unless colored sparticles are very heavy, the SUSY events at the LHC
mainly result in gluino and squark production, followed by their
possibly lengthy cascade decays. These events, therefore, typically
contain very hard jets (from the primary decay of the squark and/or
gluino) together with other jets and isolated electrons, muons and
taus (identified as narrow one- and three-prong jets), and sometimes
also photons, from the decays of secondary charginos and neutralinos,
along with $\eslt$ that arises from the escaping dark matter particles
(as well as from neutrinos).  In models with a superlight gravitino, there
may also be additional isolated photons, leptons or jets from the decay
of the NLSP.  The relative rates for various $n$-jet + $m$-lepton +
$k$-photon +$\eslt$ event topologies is sensitive to the model as well
as to the parameter values, and so provide a useful handle for
phenomenological analyses.

Within the SM, the physics background to the classic $jets+ \eslt$ signal
comes from neutrinos escaping the detector.  Thus, the dominant SM
backgrounds come from $W+jets$ and $Z+jets$ production, $t\bar{t}$
production, QCD multijet production (including $b\bar{b}$ and $c\bar{c}$
production), $WW,\ WZ,\ ZZ$ production plus a variety of $2\to n$
processes which are not usually included in event generators.  These
latter would include processes such as $t\bar{t}t\bar{t}$,
$t\bar{t}b\bar{b}$, $t\bar{t}W$, $WWW$, $WWZ$ production, {\it etc.}
Decays of electroweak gauge bosons and the $t$-quark are the main source
of isolated leptons in the SM. Various additional effects--
uninstrumented regions, energy mis-measurement, cosmic rays, 
beam-gas events-- can also lead to $\eslt$ events.  

In contrast to the SM, SUSY events naturally tend to have large jet
multiplicities and frequently an observable rate for high multiplicity lepton
events with large $\eslt$. 
Thus, if one plots signal and background versus multiplicity of any
of these quantities, as one steps out to large multiplicity, the expected 
SUSY events should increase in importance, and even dominate the 
high multiplicity channels in some cases. 
This is especially true of isolated multi-lepton
signatures, and in fact it is convenient to classify SUSY signal 
according to lepton multiplicity\cite{btw}:
\bi
\item zero lepton $+jets+\eslt$ events,
\item one lepton $+jets+\eslt$ events, 
\item two opposite sign leptons $+jets+\eslt$ events (OS), 
\bi
\item same-flavor (OSSF),
\item different flavor (OSDF),
\ei
\item two same sign leptons $+jets+\eslt$ events (SS), 
\item three leptons $+jets+\eslt$ events ($3\ell$), 
\item four (or more) leptons $+jets+\eslt$ events ($4\ell$).
\ei

\subsection{LHC reach for SUSY}

Event generators, together with detector simulation programs can be used
to project the SUSY discovery reach of the LHC. 
Given a specific model, one may first generate a
grid of points that samples the parameter (sub)space
where signal rates are expected to vary significantly. 
A large number of SUSY collider events can then be generated 
at every point on the grid along with
the various SM backgrounds to the SUSY signal mentioned above. 
Next, these signal and background events are passed through 
a detector simulation program 
and a jet-finding algorithm is implemented to determine
the number of jets per event above some $E_T(jet)$ threshold (usually
taken to be $E_T(jet)>50-100$~GeV for LHC).  Finally, 
{\it   analysis cuts} are imposed which are 
designed to reject mainly SM BG while retaining the signal.  These cuts 
may include both topological and kinematic selection criteria.
For observability with an assumed integrated luminosity, we require that
the signal exceed the chance 5 standard deviation upward fluctuation of
the background, together with a minimum value of ($\sim 25$\%) the
signal to background ratio, to allow for the fact that the background is
not perfectly known.  For lower sparticle masses, softer kinematic cuts
are used, but for high sparticle masses, the lower cross sections but
higher energy release demand hard cuts to optimize signal over
background.

In Fig.~\ref{fig:lhc}, we illustrate the SUSY reach of the LHC within
the mSUGRA model assuming an integrated luminosity of
100~fb$^{-1}$. We show the result in the $m_0-m_{1/2}$ plane, taking
$A_0=0$, $\tan\beta =10$ and $\mu >0$.  The signal is observable over
background in the corresponding topology below the
corresponding curve.  We note the following.

\begin{enumerate}
\item Unless sparticles are very heavy, there is  an observable  signal 
in several different event topologies. This will help add confidence
that one is actually seeing new physics, and may help to sort out the
production and decay mechanisms. 
\item The reach at low $m_0$ extends to $m_{1/2}\sim 1400$~GeV. This
corresponds to a reach for $m_{\tq}\sim m_{\tg}\sim 3.1$ TeV.
\item At large $m_0$, squarks and sleptons are in the $4-5$ TeV range,
and are too heavy to be produced at significant rates at LHC. Here, the reach
comes mainly from just gluino pair production. In this range, the LHC
reach is up to $m_{1/2}\sim 700$ GeV, corresponding to a reach in
$m_{\tg}$ of about 1.8 TeV, and may be extended by $\sim$ 15-20\% by
$b$-jet tagging\cite{mizukoshi}. 
\end{enumerate}
\begin{figure}[tbh]
\begin{center}
\includegraphics[width=7cm]{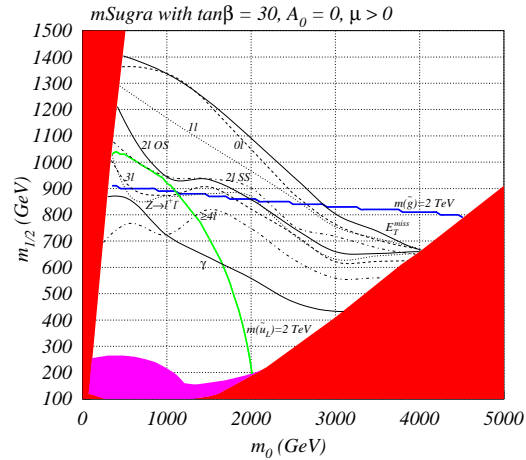}
\end{center}
\vspace{-3mm}
\caption{\small\it 
The 100 fb$^{-1}$ fb reach of LHC for SUSY in the mSUGRA model. For each
event 
topology, the signal is observable below the corresponding contour. }
\label{fig:lhc} 
\end{figure}

In Fig. \ref{fig:sugra} we can see a comparison of the LHC reach
(notice that it is insensitive to $\tan\beta$ and sign($\mu$)) with 
that of the Tevatron (for clean $3\ell$
events with 10 fb$^{-1}$), and the proposed
$e^+e^-$ International Linear Collider (ILC),   with $\sqrt{s}=0.5$ or 1 TeV
along with various dark matter direct detection (DD)  and 
indirect detection (ID) search experiments. 
We remark that:
\bi
\item While LHC can cover most of the relic density allowed region, the HB/FP
region emerges far beyond the LHC reach. 
\item As will be discussed, the DD and ID experiments have the greatest
sensitivity in the HB/FP region where the neutralino is MHDM. 
In this sense, DD and ID experiments {\it complement} LHC searches for SUSY.
\item The ILC reach is everywhere lower than LHC, except in the HB/FP
region.  In this region, while gluinos and squarks can be extremely
heavy, the $\mu$ parameter is small, leading to a relatively light
spectrum of charginos and neutralinos. These are not detectable at the
LHC because the visible decay products are too soft.  However, since
chargino pair production is detectable at ILC even if the energy release
in chargino decays is small, the ILC reach extends beyond LHC in this
region\cite{tadas}.  
\ei
Finally, we note here that while the results presented above are 
for the LHC reach in the mSUGRA model, 
the LHC reach (measured in terms of $m_{\tg}$ and
$m_{\tq}$) tends to be relatively insensitive to the 
details of the model chosen, as long as gluino and squark production
followed by cascade decays to the DM particle occur. 

\begin{figure}[tbh]
\begin{center}
\includegraphics[width=7cm]{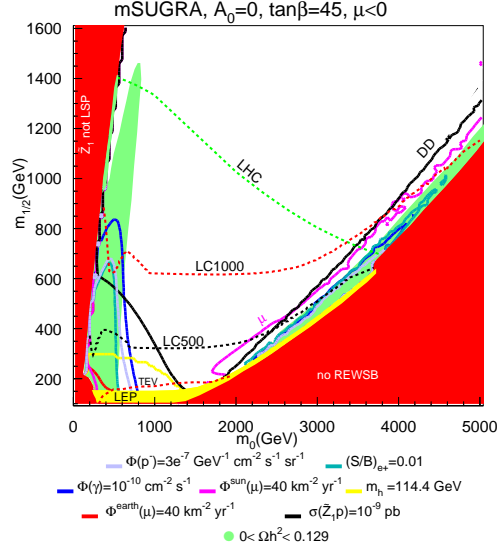}
\end{center}
\vspace{-3mm}
\caption{\small\it The projected reach of various colliders, direct and
  indirect dark matter search experiments in the mSUGRA model. For the
  indirect search results we have adopted the 
  conservative default DarkSUSY isotropic
  DM halo density distribution. Plot is from Ref. \cite{bbko}. }
\label{fig:sugra} 
\end{figure}

\subsection{Early discovery of SUSY at LHC without $\eslt$}

The classic signature for SUSY at hadron colliders is the presence of
jets, isolated leptons and especially large $\eslt$. However, in the early period 
of LHC running, $\eslt$ may be very difficult to measure reliably. In a real detector, 
there will be calorimetry calibration issues, dead or uninstrumented regions, 
``hot'' cells, cosmic ray events, beam-gas collisions: all of these contribute
to $\eslt$ events, in additon to SM backgrounds and possibly new physics.
As learned from Tevatron experiments, a reliable measurement of $\eslt$ requires
knowledge of the complete detector, and that may well take some time--
possibly over a year-- after start-up. 

Can LHC search for SUSY even if $\eslt$ searches are not viable? 
The answer seems to be yes, as long as lepton isolation cuts are possible.
In Ref. \cite{etmiss}, it was shown that requiring events with $\ge 4$ jets plus
a {\it high multiplicity} of isolated leptons: SS, OSSF, OSDF, $3\ell$, $\cdots$
would allow one to severely reduce SM background while maintaining
large enough signal rates. It is claimed that
an LHC reach of $m_{\tg}\sim 750$ GeV is possible with just $0.1$ fb$^{-1}$ of integrated
luminosity, {\it without} using an $\eslt$ cut, by requiring events with $\ge$ 3 isolated
leptons ($e$s or $\mu$s). 

In addition, reliable electron identification may also be an issue in early LHC running.
The problem here is differentiating the electron's EM shower from other energy
depositions such as charged pions, or hard photons where in additon a charged track is nearby.
In this case, a SUSY search may still be made by looking for multi-jet plus
{\it multi-muon} production, again without any requirement on missing $E_T$\cite{mm}.
Atlas and CMS are already reliably detecting cosmic muons. Also, muons enjoy an 
addvantage over electrons in that they can be detected reliably to $p_T$
values as low as 5 GeV.

The multiplicity of muons in $\ge 4$ jet events is shown in Fig. \ref{fig:multimu},
assuming $p_T(\mu )>10$ GeV to remove BG from heavy flavor decays in SM processes.
Here, we see that as one moves to high muon multiplicity, the SM background drops off
sharply, and by $n_\mu \ge 3$, signal already far exceeds BG, at least for the
SPS1a' signal point shown in the figure. In fact, by requiring {\it same-sign dimuon}
plus multi-jet events, already signal in many cases will exceed background.
The LHC should be able to discover
gluinos with $m_{\tg}\sim 450$ ($550$) GeV in the SS dimuon plus $\ge 4$ jets
state with just 0.04 (0.1) fb$^{-1}$ of integrated luminosity.
\begin{figure}[tbh]
\begin{center}
\includegraphics[width=8cm]{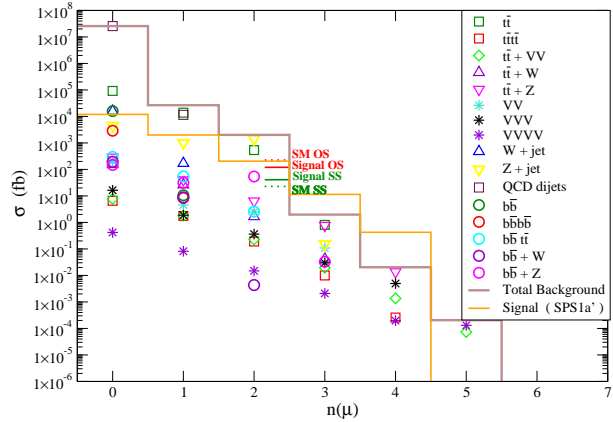}
\end{center}
\vspace{-3mm}
\caption{\small\it 
Multiplicity of isolated muons in $\ge 4$ jets events, from signal point
SPS1a' and various SM background processes.
}
\label{fig:multimu} 
\end{figure}

\subsection{Determination of sparticle properties}
\label{subsec:properties}

Once a putative signal for new physics emerges at LHC, the next
step is to establish its origin. This will entail detailed measurements of cross
sections and distributions in various event topologies to gain insight
into the identity of the new particles being produced, their masses, decay
patterns, spins, couplings (gauge quantum numbers) and ultimately mixing
angles. These measurements are not straightforward in the LHC environment
because of numerous possible SUSY production reactions occurring simultaneously, 
a plethora of sparticle cascade decay possibilities,
hadronic debris from initial state radiation and lack of invariant mass
reconstruction due to the presence of $\eslt$. All these lead to
ambiguities and combinatoric problems in reconstructing exactly what
sort of signal reactions are taking place.
In contrast, at the ILC, the initial state is simple, the
beam energy is tunable and beam polarization can be used to select out
specific processes.

While it seems clear that the ILC is better suited for a systematic
program of precision sparticle measurements, studies have shown (albeit
in special cases) that interesting measurements are also possible at the
LHC. We go into just a subset of all details here in order to give the
reader an idea of some of the possibilities suggested in the literature.

One suggested starting point is the distribution of effective mass
$M_{\rm eff}= \eslt +\sum E_T(jets)$ in the inclusive SUSY sample, which
sets the approximate mass scale $M_{\rm SUSY}\equiv {\rm min}(m_{\tg},
m_{\tq})$ for the strongly interacting sparticles are being
produced\cite{frank1}, and provides a measure of $M_{\rm SUSY}$ to
10-15\%. 

More detailed information on sparticle masses may be accessed
by studying specific event topologies. For instance,  the mass of
dileptons from $\tz_2 \to \ell^+\ell^-\tz_1$ decays is bounded by
$m_{\tz_2}-m_{\tz_1}$ (this bound is even more restrictive if $\tz_2$
decays via an on-shell slepton)\cite{mlledge}.
We therefore expect an OSSF invariant mass  distribution to exhibit an edge 
at $m_{\tz_2}-m_{\tz_1}$ (or below) in any sample of SUSY events
so long as the ``spoiler'' decay modes $\tz_2\to\tz_1 Z$ or $\tz_1 h$ 
are closed.
Contamination from chargino production can be statistically
removed by subtracting out the distribution of OSDF dileptons. 
In MHDM models, there may be more than one visible mass edge because
the $\tz_3$ may also be accessible in cascade decays.

In the happy circumstance where production of gluinos or a 
single type of squark is dominant, 
followed by a string of two-body decays, then further invariant mass edges
are possible. One well-studied example comes from $\tg\to
b\bar{\tb}_1 \to b\bar{b}\tz_2\to b\bar{b}\ell\bar{\ell}\tz_1$; then one
can try to combine a $b$-jet with the dilepton pair to reconstruct the
squark-neutralino mass edge:
$m(b\ell\bar{\ell})<m_{\tb_1}-m_{\tz_1}$. Next, combining with
another $b$-jet can yield a gluino-neutralino edge:
$m(b\bar{b}\ell\bar{\ell})<m_{\tg}-m_{\tz_1}$.  The reconstruction of
such a decay chain may be possible as shown in Ref.~\cite{frank1}, where
other sequences of two-body decays are also examined.  In practice, such
fortuitous circumstances may not exist, and there are many combinatoric
issues to overcome as well.  A different study\cite{dutta} shows that
end-point measurements at the LHC will make it possible to access the
mass difference between the LSP and the stau in a mSUGRA scenario where
the stau co-annihilation mechanism is operative.

These end-point measurements generally give mass differences, not
masses. However, by an analysis of the decay chain $\tq_L \to q\tz_2 \to        
q\tell^\pm\ell^\mp \to q\ell^\pm\ell^\mp\tz_1$, it has been
argued\cite{frank2} that reconstruction of masses may be possible under
fortuituous circumstances. More recently, it has been suggested that it
may be possible to directly access the gluino and/or squark masses (not
mass differences) via the introduction of the so-called $m_{T2}$
variable. We will refer the reader to the literature for
details\cite{mt2}. It also may be possible to make a measurement of $m_{\tg}$
basd on total production cross section in cases where pure gluino pair production
is dominant\cite{gabe}.

Mass measurements allow us to check consistency of specific SUSY models
with a handful of parameters, and together with other measurements can
readily exclude such models. But these are not the only interesting
measurements at the LHC. It has been shown that if the NLSP of GMSB
models decays into a superlight gravitino, it may be possible to
determine its lifetime, and hence the gravitino mass at the
LHC\cite{lifetime}. This will then allow one to infer the underlying SUSY
breaking scale, a scale at least as important as the weak scale! A
recent study\cite{yanagida} suggests that this is possible even when the
the decay length of the NLSP is too short to be measured. While linear
collider experiments will ultimately allow the precision measurements
that will directly determine the new physics to be softly broken
supersymmetry\cite{peskin}, it will be exciting to analyze real LHC data
that will soon be available to unravel many of the specific details
about how (or if) SUSY is actually implemented in nature.

\subsection{Measuring DM properties at LHC and ILC}

SUSY discovery will undoubtedly be followed by a program (as outlined in
Sec.~\ref{subsec:properties}) to reconstruct sparticle properties.  What
will we be able to say about dark matter in light of these measurements?
Such a study was made by Baltz {\it et al.}\cite{bbpw} where four
mSUGRA case study points (one each in the bulk region, the HB/FP region,
the stau coanihilation region and the $A$-funnel region) were examined
for the precision with which measurements of sparticle properties that
could be made at LHC, and also at a $\sqrt{s}=0.5$ and 1 TeV $e^+e^-$
collider.  They then adopted a 24-parameter version of the MSSM and fit
its parameters to these projected measurements. 
The model was then used to predict several quantities
relevant to astrophysics and cosmology: the dark matter relic density
$\Omega_{\tz_1}h^2$, the spin-independent neutralino-nucleon scattering
cross section $\sigma_{SI}(\tz_1 p)$, and the neutralino annihilation
cross section times relative velocity, in the limit that $v\to 0$:
$\langle\sigma v\rangle |_{v\to 0}$.  The last quantity is the crucial
particle physics input for estimating signal strength from neutralino
annihilation to anti-matter or gammas in the galactic halo. 
What this yields then is a {\it collider measurement} of these
key dark matter quantities.

As an illustration, we show in Fig. \ref{fig:lcc2} (taken from
Ref.~\cite{bbpw}) the precision with which the neutralino relic density
is constrained by collider measurements for the LCC2 point which is in
the HB/FP region of the mSUGRA model. Measurements at the LHC cannot fix
the LSP composition, and so are unable to resolve the degeneracy between a
wino-LSP solution (which gives a tiny relic density) and the true
solution with MHDM. Determinations of chargino production cross sections
at the ILC can easily resolve the difference. It is nonetheless striking
that up to this degeneracy ambiguity, experiments at the LHC can pin down the 
relic density to within $\sim 50$\% (a remarkable result, given that 
there are sensible models where the predicted relic density may
differ  by orders of
magnitude!). This improves to 10-20\% if we can combine LHC and
ILC measurements. 
\begin{figure}[tbh]
\begin{center}
\includegraphics[width=7cm]{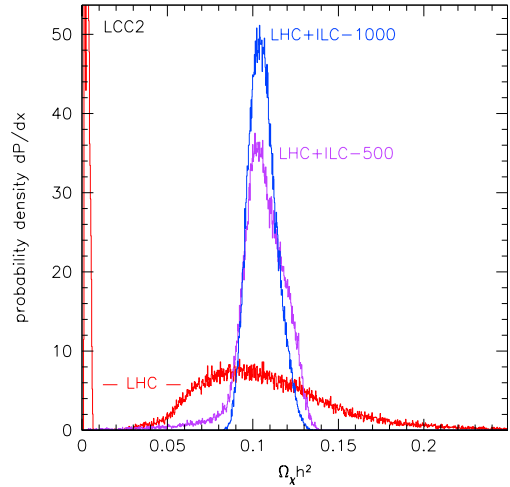}
\end{center}
\vspace{-3mm} 
\caption{\small\it 
Determination  of neutralino relic abundance via measurements
at the LHC and ILC, taken  from Ref.~\cite{bbpw}.}
\label{fig:lcc2} 
\end{figure}

This collider determination of the relic density is very important. If
it agrees with the cosmological measurement it would establish that the
DM is dominantly thermal neutralinos from the Big Bang.  If the
neutralino relic density from colliders falls significantly below
(\ref{eq:wmap}), it would provide direct evidence for multi-component DM--
perhaps neutralinos plus axions or other exotica. Alternatively, if the
collider determination gives a much larger value of $\Omega_{\tz_1}h^2$,
it could point to a long-lived but unstable neutralino and/or
non-thermal DM.

The collider determination of model parameters would also pin down the
neutralino-nucleon scattering cross section. Then if a WIMP signal is actually
observed in DD experiments, one might be able to determine the local DM
density of neutralinos and aspects of their velocity distribution based on the DD
signal rate. This density should agree with that obtained from
astrophysics if the DM in our Galaxy is comprised only of neutralinos. 

Finally, a collider determination of $\langle\sigma v\rangle |_{v\to 0}$
would eliminate uncertainty on the particle physics side of projections for
any ID signal from annihilation of neutralinos in the galactic halo.
Thus, the observation of a gamma ray and/or anti-matter signal from 
neutralino halo annihilations would facilitate the determination
of the galactic halo dark matter density distribution. 

\subsection{Some non-SUSY WIMPs at the LHC} \label{sec:other}

\subsubsection{$B_{\mu}^1$ state from universal extra dimensions}

Models with Universal Extra Dimensions, or UED, are interesting
constructs which provide a foil for SUSY search analyses\cite{KKreview}.
In the 5-D UED theory, one posits that the fields of the SM actually
live in a 5-D brane world. The extra dimension is ``universal'' since
{\it all} the SM particles propagate in the 5-D bulk. The single extra
dimension is assumed to be compactified on a $S_1/Z_2$ orbifold (line
segment).  After compactification, the 4-D effective theory includes the
usual SM particles, together with an infinite tower of Kaluza-Klein (KK)
excitations. The masses of the excitations depend on the radius of the
compactified dimension, and the first ($n=1$) KK excitations can be
taken to be of order the weak scale.  In these theories, KK-parity
$(-1)^n$ can be a conserved quantum number. If this so-called KK-parity
is exact, then the lightest odd KK parity state will be stable and 
can be a DM candidate.
At tree-level, all the KK excitations in a given level are essentially
degenerate. 
Radiative corrections break the degeneracy,
leaving colored excitations as the heaviest excited states and 
the $n=1$ KK excitation of the SM $U(1)_Y$ gauge boson $B_{\mu}^1$
as the lightest\cite{matchev} KK odd state: in the UED case, therefore,
the DM particle has spin-1.
The splitting caused by the
radiative corrections is also essential to assess how the KK excitations
decay, and hence are crucial for collider phenomenology \cite{rizzo}.

The relic density of $B_\mu^1$ particles has been computed, and found to
be compatible with observation for certain mass ranges of
$B_\mu^1$\cite{uedrelic}. Also, in UED, the colored excitations can be
produced with large cross sections at the LHC, and decay via a cascade
to the $B_\mu^1$ final state. Thus, the collider signatures are somewhat
reminiscent of SUSY, and it is interesting to ask whether it is possible
to distinguish a $jets+leptons+\eslt$ signal in UED from that in SUSY.
Several studies\cite{kong} answer affirmatively, and in fact provide
strong motivation for the measurement of the spins of the produced new
particles\cite{spin}. UED DM generally leads to a large rate in IceCube,
and may also give an observable signal in anti-protons and possibly also
in photons and
positrons\cite{KKreview,bnps}. DD is also possible but the SI cross
section is typically smaller than $10^{-9}$~pb.

\subsubsection{Little Higgs models}
\label{sec:lh}

Little Higgs models \cite{lhorig,LHreview} provide an alternative method
compared to SUSY 
to evade the quadratic sensitivity of the scalar Higgs sector to
ultra-violet (UV) physics. In this framework, the Higgs boson is a
pseudo-Goldstone boson of a spontaneously broken global symmetry that is
not completely broken by any one coupling, but is broken when 
{\it all} couplings are included. This then implies that
there is quadratic sensitivity to UV physics, but only at the multi-loop
level. Specific models where the quadratic sensitivity enters at the
two-loop level should, therefore, be regarded as low energy
effective theories valid up to a scale $\Lambda \sim 10$~TeV, at which a
currently unknown, and perhaps strongly-coupled UV completion of the
theory is assumed to exist. Models that realize this idea require
new TeV-scale degrees of freedom that can be searched for at the LHC:
new gauge bosons, a heavy top-like quark, and new spin-zero particles, all
with couplings to the SM. These models, however, run into
phenomenological difficulties with precision EW constraints, unless a
discrete symmetry-- dubbed $T$-parity\cite{cl}-- is included.  
SM particles are then $T$-even, while the new particles are $T$-odd.

We will set aside the issue of whether $T$-parity
conservation is violated by anomalies\cite{hill}, and assume that a
conserved $T$-parity can be introduced\cite{uvcomp}.  In this case, the
lightest $T$-odd particle $A_H$ -- the Little Higgs partner of the
hypercharge gauge boson with a small admixture of the neutral $W_{3H}$
boson -- is stable and yields the observed amount of DM for a reasonable
range of model parameters\cite{bnps}. In this case, the DM particle has
spin-1, though other cases with either a spin-$\frac{1}{2}$ or spin-0
heavy particle may also be possible. $A_H$ can either annihilate with
itself into vector boson pairs or $t\bar{t}$ pairs via $s$-channel Higgs
exchange, or into top pairs via exchange of the heavy $T$-odd quark in
the $t$-channel. Co-annihilation may also be possible if the heavy quark
and $A_H$ are sufficiently close in mass. Signals at the LHC\cite{ccy,bel}
mainly come from pair production of heavy quarks $pp\to T\bar{T}X$ followed by
$T\to t A_H$ decay, and from single
production of the heavy quark in association with $A_H$. These lead to
$t\bar{t}+\eslt$ events at LHC\cite{tao}. The $\eslt$ comes from the
escaping $A_H$ particle, which must be the endpoint of all $T$-odd
particle decays.\footnote{We note here that it is also possible to
construct so-called twin-Higgs models\cite{cgh} where the Higgs sector
is stabilized via new particles that couple to the SM Higgs doublet, but
are singlets under the SM gauge group. In this case, there would be no
obvious new physics signals at the LHC.}  If $A_H$ is the dominant
component of galactic DM, we will generally expect small DD and ID rates
for much the same reasons that the signals from the bino LSP tend to be
small\cite{bnps}: see, however, Ref.\cite{bai} for a different model
with large direct detection rate.

\section{Lecture 4: $\eslt$ and the dark matter connection}
\label{sec:lec4}

\subsection{Neutralino relic density}

Once a SUSY model is specified, then given a set of input parameters, it
is possible to compute all superpartner masses and couplings necessary
for phenomenology.  We can then use these to calculate scattering cross
sections and sparticle decay patterns to evaluate SUSY signals (and
corresponding SM backgrounds) in collider experiments.  We can also
check whether the model is allowed or excluded by experimental
constraints, either from direct SUSY searches, {\it e.g.} at LEP2 which
requires that $m_{\tw_1}>103.5 $ GeV, $m_{\te}\agt 100$~GeV, and
$m_h>114.4$ GeV (for a SM-like light SUSY Higgs boson $h$), or from
indirect searches through loop effects from SUSY particles in low energy
measurements such as $B(b\to s\gamma)$ or $(g-2)_\mu$.  We can also
calculate the expected thermal LSP relic density.  To begin our
discussion, we will first assume that the lightest neutralino $\tz_1$ is
the candidate DM particle.

As mentioned earlier, the relic density calculation involves solving the
Boltzmann equation, where the neutralino density changes due to both
the expansion of the Universe and because of neutralino annihilation into
SM particles, determined by the thermally averaged $\tz_1\tz_1$
annihilation cross section. An added complication occurs if neutralino
{\it co-annihilation} is possible. Co-annihilation occurs if there is
another SUSY particle close in mass to the $\tz_1$, whose thermal relic
density (usually suppressed by the Boltzmann factor $exp{-\Delta M\over
T}$) is also significant. In the mSUGRA model, co-annihilation may occur
from a stau, $\ttau_1$, a stop $\tst_1$ or the lighter chargino $\tw_1$.
For instance, in some mSUGRA parameter-space regions the $\ttau_1$ and
$\tz_1$ are almost degenerate, so that they both have a significant density
in the early universe,
and reactions such as $\tz_1\ttau_1\to \tau\gamma$ occur. Since the 
electrically charged $\ttau_1$ can also annihilate efficiently via
electromagnetic interactions, this process also alters the equilibrium
density of
neutralinos.  All in all, there are well over a thousand neutralino
annihilation and co-annihilation reactions that need to be computed,
involving of order 7000 Feynman diagrams. There exist several publicly
available computer codes that compute the neutralino relic density: these
include DarkSUSY\cite{dsusy}, MicroMegas\cite{micro} and
IsaReD\cite{isared} (a part of the Isatools package of
Isajet\cite{isajet}).

As an example, we show in Fig. \ref{fig:pspace} the $m_0\ vs.\ m_{1/2}$
plane from the mSUGRA model, where we take $A_0=0$, $\mu >0$,
$m_t=171.4$ GeV and $\tan\beta =10$.  The red-shaded regions are
not allowed because either the $\ttau_1$ becomes the lightest SUSY
particle, in contradiction to negative searches for long lived, charged
relics (left edge), or EWSB is not correctly obtained (lower-right
region). The blue-shaded region is excluded by LEP2 searches
for chargino pair production ($m_{\tw_1}<103.5$ GeV).
We
show contours of squark (solid) and gluino (dashed) mass
(which are nearly invariant under change of $A_0$ and $\tan\beta$).
Below the magenta contour near $m_{1/2}\sim 200$~GeV, 
$m_h<110$ GeV, 
which is roughly the
LEP2 lower limit on $m_h$ in the model. 
The
thin green regions at the edge of the unshaded white
region has $0.094< \Omega_{\tz_1}h^2< 0.129$ where the neutralino
saturates the observed relic density.  In the adjoining
yellow regions, $\Omega_{\tz_1}h^2<0.094$, so these
regions require multiple DM components. The white regions all have
$\Omega_{\tz_1}h^2>0.129$ and so give too much thermal DM: they are
excluded in the standard Big Bang cosmology.
\begin{figure}[hbt]
\begin{center}
\includegraphics[width=7cm]{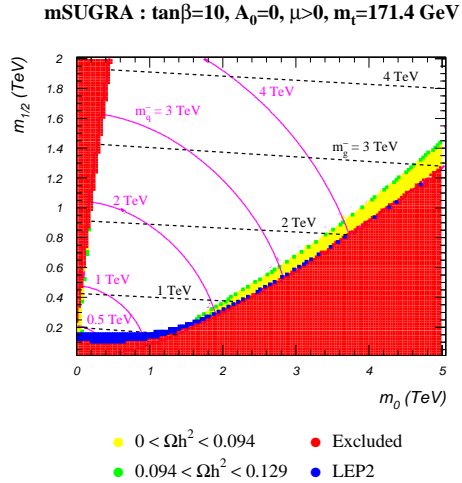}\\
\end{center}
\vspace{-3mm}
\caption{\small\it DM-allowed regions in the $m_0-m_{1/2}$ plane of the
mSUGRA model for $\tan\beta =10$ with $A_0=0$ and $\mu>0$. }
\label{fig:pspace} 
\end{figure}

The DM-allowed regions are classified as follows:
\bi
\item At very low $m_0$ and low $m_{1/2}$ values is the so-called {\it
bulk} annihilation region\cite{bulk}.  Here, sleptons are quite
light, so $\tz_1\tz_1\to \ell\bar{\ell} $ via $t$-channel slepton
exchange. In years past (when $\Omega_{\rm CDM}h^2 \sim 0.3$ was quite
consistent with data), this was regarded as the favored region.  But
today LEP2 sparticle search limits have increased the LEP2-forbidden
region from below, while the stringent bound $\Omega_{CDM}h^2\leq 0.13$
has pushed the DM-allowed region down. Now hardly any bulk region
survives in the mSUGRA model.

\item At low $m_0$ and moderate $m_{1/2}$, there is a thin strip of
(barely discernable) allowed region adjacent to the stau-LSP region where
the neutralino and the lighter stau were in thermal equilibrium in the
early universe. Here co-annihilation with the light stau serves to bring
the neutralino relic density down to its observed value\cite{stau}. 
\item At large $m_0$, adjacent to the EWSB excluded region on the right, is the
hyperbolic branch/focus point (HB/FP) region, where the superpotential
$\mu$ parameter becomes small and the higgsino-content of $\tz_1$
increases significantly. Then $\tz_1$ can annihilate efficiently via
gauge coupling to its higgsino component and becomes mixed
higgsino-bino DM. If $m_{\tz_1}>M_W,\ M_Z$, then
$\tz_1\tz_1\to WW,\ ZZ,\ Zh$ is enhanced, and one finds the correct
measured relic density\cite{hb_fp}.
\ei

We show the corresponding situation for $\tan\beta=52$ in
 Fig. \ref{fig:pspace52}. While the stau co-annihilation and the HB/FP
 regions are clearly visible, we see that a large DM consistent
 region now appears. 
\bi
\item In this region, the value of $m_A$ is small enough
 so that $\tz_1\tz_1$ can annihilate into $b\bar{b}$ pairs through
 $s$-channel $A$ (and also $H$) resonance.  This
 region has been dubbed the $A$-funnel\cite{Afunnel}. 
It can be very broad at large
 $\tan\beta$ because the width $\Gamma_A$ can be quite wide due to 
the very large $b$- and $\tau$- Yukawa
 couplings.
If $\tan\beta$ is increased further, then $\tz_1\tz_1$ annihilation 
through the (virtual) $A^*$ is large all over parameter space, 
and most of the theoretically-allowed
parameter space becomes DM-consistent. For even higher $\tan\beta$ 
values, the parameter space collapses due to a lack of 
appropriate EWSB.
\ei
\begin{figure}[hbt]
\begin{center}
\includegraphics[width=7cm]{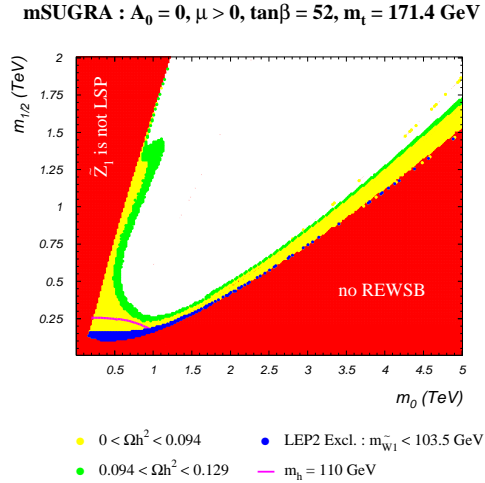}\\
\end{center}
\vspace{-3mm}
\caption{\small\it DM-allowed regions in the $m_0-m_{1/2}$ plane of the
mSUGRA model for $\tan\beta =52$ with $A_0=0$ and $\mu>0$. The various
colors of shading is as in Fig.~\ref{fig:pspace}.}
\label{fig:pspace52} 
\end{figure}

\vspace{-1mm}
It is also possible at low $m_{1/2}$ values that a
light Higgs $h$ resonance annihilation region can occur just above the
LEP2 excluded region\cite{hfunnel}.  Finally, if $A_0$ is large and negative,
then the $\tst_1$ can become light, and  $m_{\tst_1}\sim m_{\tz_1}$, 
so that stop-neutralino co-annihilation\cite{stop_co} can occur.

Up to now, we have confined our discussion to the mSUGRA framework in
which compatibility with (\ref{eq:wmap}) is obtained only over selected
portions of the $m_0-m_{1/2}$ plane. The reader may well wonder what
happens if we relax the untested universality assumptions that underlie
mSUGRA. Without going into details, we only mention here that in many
simple one-parameter extensions of mSUGRA where the universality of mass
parameters is relaxed in any one of the matter scalar, the Higgs scalar,
or the gaugino sectors, {\it all points in the $m_0-m_{1/2}$ plane
become compatible with the relic density constraint} due to a variety of
mechanisms: these are catalogued in Ref. \cite{wtnreview}.  Implications
of the relic density measurement for collider searches must thus be drawn
with care.

\subsection{Neutralino direct detection}

Fits to galactic rotation curves imply a local
relic density of $\rho_{CDM}\sim 0.3$ GeV/cm$^3$. For a 100 GeV WIMP,
this translates to about one WIMP per coffee mug volume at our location
in the galaxy. The goal of DD experiments is to detect the very rare
WIMP-nucleus collisions that should be occuring as the earth, together
with the WIMP detector, moves through the DM halo.

DD experiments are usually located deep underground to shield the
experimental apparatus from background due to cosmic rays and ambient
radiation from the environment or from radioactivity induced by cosmic
ray exposure. One technique is to use cryogenic crystals cooled to near
absolute zero, and look for phonon and ionization signals from nuclei
recoiling from a WIMP collision.  In the case of the CDMS
experiment\cite{cdms} at the Soudan iron mine, target materials include
germanium and silicon. Another technique uses noble gases cooled to a
liquid state as the target. Here, the signal is scintillation light
picked up by photomultiplier tubes and ionization.  Target materials
include xenon\cite{xenon10}, argon and perhaps neon. These noble liquid
detectors can be scaled up to large volumes at relatively low cost. They
have the advantage of fiducialization, wherein the outer layers of the
detector act as an active veto against cosmic rays or neutrons coming
from phototubes or detector walls: only single scatters from the inner
fiducial volume qualify as signal events.  A third technique, typified
by the COUPP experiment\cite{coupp}, involves use of superheated liquids
such as $CF^3I$ located in a transparent vessel. The nuclear recoil from
a WIMP-nucleon collision then serves as a nucleation site, so that a
bubble forms. The vessel is monitored visually by cameras. Background
events are typically located close to the vessel wall, while neutron
interactions are likely to cause several bubbles to form, instead of
just one, as in a WIMP collision.  This technique allows for the use of
various target liquids, including those containing elements such as
fluorine, which is sensitive to {\it spin-dependent} interactions.

The cross section for WIMP-nucleon collisions can be calculated, and in
the low velocity limit separates into a coherent spin-independent
component (from scattering mediated by scalar quarks and scalar Higgs
bosons) which scales as nuclear mass squared, and a spin-dependent
component from scattering mediated by the $Z$ boson or by squarks,
which depends on the WIMP and nuclear spins\cite{dgr}. The scattering
cross section per nucleon versus $m_{WIMP}$ 
serves as a figure of merit and facilitates
the comparison of the sensitivity of various experiments using different
target materials.

\begin{figure}[hbt]
\begin{center}
\includegraphics[width=8cm]{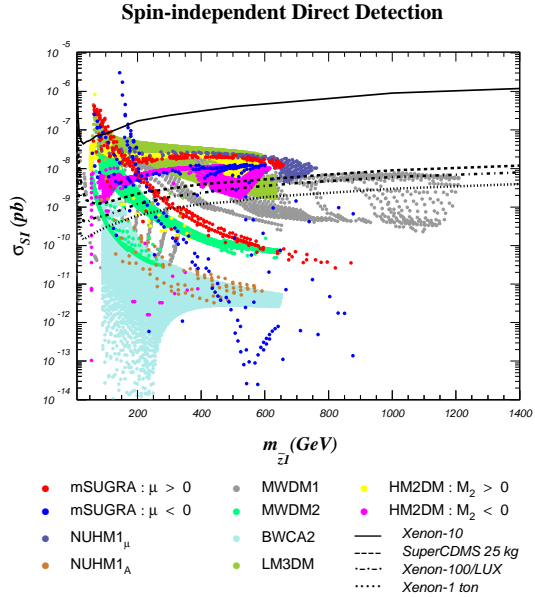}\\
\end{center}
\vspace{-3mm}
\caption{\small\it 
The spin-independent neutralino-proton scattering cross-section vs
$m_{\tz_1}$ in a variety of SUSY models, compatible
with collider constraints where thermally produced  Big Bang neutralinos
saturate the observed dark matter density.}
\label{fig:dd} 
\end{figure}
In Fig. \ref{fig:dd}, we show the spin-independent
$\tz_1 p$ cross section versus $m_{\tz_1}$ for a large number of SUSY
models (including mSUGRA). Every color represents a different
model. For each model, parameters are chosen so that current collider
constraints on sparticle masses are satisfied, and further, that the
lightest neutralino (assumed to be the LSP) saturates the observed relic
abundance of CDM.  Also shown is the sensitivity of current experiments
together with projected sensitivity of proposed searches at superCDMS,
Xenon-100, LUX, WARP and at a ton-sized noble liquid detector.
The details of the various models are unimportant for our present
purpose. The key thing to note is that while the various models have a
branch where $\sigma_{\rm SI}(p\tz_1)$ falls off with $m_{\tz_1}$, there
is another branch where this cross-section asymptotes to just under
$10^{-8}$~pb\cite{wtn,wtnreview,pran}. 
Points in thisbranch (which includes the HB/FP region of
mSUGRA), are consistent with (\ref{eq:wmap}) because $\tz_1$ has a
significant higgsino component.  Neutralinos with an enhanced higgsino
content can annihilate efficiently in the early universe via gauge
interactions. Moreover, since the spin-independent DD amplitude is
mostly determined by the Higgs boson-higgsino-gaugino coupling, it is
large in models with MHDM which has both gaugino and higgsino
components. Thus the enhanced higgsino component of MHDM increases both
the neutralino
annihilation in the early universe as well as the spin-independent DD
rate. The exciting thing is that the experiments currently being deployed-- 
such as Xenon-100, LUX and WARP-- 
will have the sensitivity to probe this class of
models. To go further will require ton-size or greater target material.

We note here that if $m_{\rm WIMP}\alt 150$ GeV, then it may be possible
to extract the WIMP mass by measuring the energy spectrum of the
recoiling nuclear targets\cite{green}.  Typically, of order 100 or more
events are needed for such a determination to 10-20\%. For higher WIMP
masses, the recoil energy spectrum varies little, and WIMP mass
extraction is much more difficult.  Since the energy transfer from the
WIMP to a nucleus is maximized when the two have the same mass, DD
experiments with several target nuclei ranging over a wide range of
masses would facilitate the distinction between somewhat light and
relatively heavy WIMPs, and so, potentially serve to establish the
existence of multiple
WIMP components in our halo.

\subsection{Indirect detection of neutralinos}

There are also a number of indirect
WIMP search techniques that attempt to detect the decay products 
from WIMP annihilation at either the center of the sun, at
the galactic center, or within the galactic halo. 

\subsubsection{Neutrino telescopes}

Neutrino telescopes such as ANTARES or IceCube can search for high energy
neutrinos produced from WIMP-WIMP annihilation into SM particles
in the core of the sun (or possibly the earth).
The technique involves detection of
multi-tens of GeV muons produced by $\nu_\mu$ interactions with polar ice
(IceCube) or ocean water (ANTARES). The muons travel at speeds greater
than the speed of light in the medium, thus leaving a tell-tale signal of
Cerenkov light which is picked up by arrays of phototubes.  The IceCube
experiment, currently being deployed at the south pole, will monitor a
cubic kilometer of ice in search of $\nu_\mu\to \mu$ conversions.  It
should be fully deployed by 2011.  The experiment is mainly sensitive to
muons with $E_\mu >50$ GeV.

In the case of neutralinos of SUSY, mixed higgsino dark matter (MHDM)
has a large (spin-dependent) cross-section to scatter from hydrogen
nuclei via $Z$-exchange and so is readily captured. Thus, in the HB/FP
region of mSUGRA, or in other SUSY models with MHDM, we expect observable
levels of signal exceeding 40 events/km$^2$/yr with $E_\mu >50$ GeV.  For
the mSUGRA model, the IceCube signal region is shown beneath the magenta
contour labelled $\mu$ in Fig. \ref{fig:sugra}\cite{bbko}. These results
were obtained using the Isajet-DarkSUSY interface\cite{dsusy}. Notice
that DD signals are also observable in much the same region (below the
contour labelled DD) where the neutralino is MHDM. 

\subsubsection{Anti-matter from WIMP halo annihilations }

WIMP annihilation in the galactic halo offers a different possibility
for indirect DM searches.  Halo WIMPs annihilate equally to matter
and anti-matter, so the rare presence of high energy anti-matter in
cosmic ray events -- positrons $e^+$, anti-protons $\pbar$, or even
anti-deuterons $\Dbar$ -- offer possible signatures.  Positrons
produced in WIMP annihilations must originate relatively close by, or
else they will find cosmic electrons to annihilate against, or lose
energy via bremsstrahlung. 
Anti-protons and anti-deuterons could originate further from us because,
being heavier, they are deflected less and so lose much less energy.
The expected signal rate depends on the WIMP
annihilation rate into anti-matter, the model for the propogation of the
anti-matter from its point of origin to the earth, and finally on the assumed
profile of the dark matter in the galactic halo. 
Several possible halo density profiles are shown in Fig.~\ref{fig:halo}.  
We see that while the {\it local} WIMP
density is inferred to a factor of $\sim 2$-3 (we are at about 8~kpc from
the Galactic center), the DM density at the galactic center is highly
model-dependent close to the core. 
Since the ID signal should scale as the
square of the WIMP density at the source, positron signals will be
uncertain by a factor of a few with somewhat larger uncertainty for
$\bar{p}$ and $\overline{D}$ signals that originate further away. 
Anti-particle propagation through the not so well known 
magnetic field leads to an additional uncertainty in the predictions.
The recently launched Pamela space-based anti-matter
telescope can look for $e^+$ or $\pbar$ events while the
balloon-borne GAPS experiment will be designed to search for
anti-deuterons. 
Anti-matter signals tend to be largest 
in the case of SUSY models with MHDM or when neutralinos annihilate
through the $A$-resonance\cite{bo}. 

\begin{figure}[tbh]
\begin{center}
\vspace{-10mm}
\includegraphics[width=7cm,angle=-90]{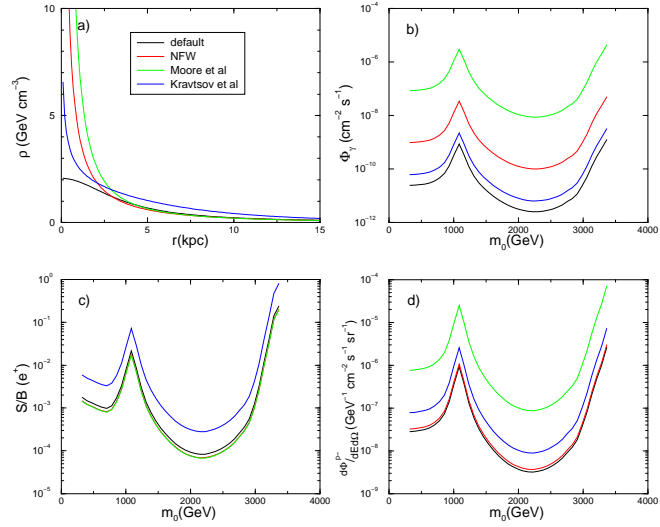}
\end{center}
\vspace{-3mm}
\caption{\small\it 
Various predictions for the DM halo in the Milky Way as a function of
distance from the galactic center. The earth is located at $r\sim 8$~kpc.}
\label{fig:halo} 
\end{figure}

\subsubsection{Gamma rays from WIMP halo annihilations}

As mentioned in the Introduction, high energy gamma rays from WIMP
annihilation offer some advantages over the signal from charged
antiparticles. Gamma rays would point to the source, and would degrade
much less in energy during their journey to us. This offers the
possibility of the line signal from $\tz_1\tz_1\to \gamma\gamma$
processes that occur via box and triangle diagrams.  While this reaction
is loop-suppressed, it yields monoenergetic photons with $E_\gamma\simeq
m_{\rm WIMP}$, and so can provide a measure of the WIMP mass. Another
possibility is to look for continuum gamma rays from WIMP annihilation
to hadrons where, for instance, the gamma is the result of $\pi^0$
decays.  Since the halo WIMPS are essentially at rest, we expect a
diffuse spectrum of gamma rays, but with $E_\gamma <m_{\rm WIMP}$.
Because gamma rays can traverse large distances, a good place to look at
is the galactic center, where the WIMP density (see Fig.~\ref{fig:halo})
is expected to be very high. Unfortunately, the density at the core is
also very uncertain, making predictions for the gamma ray flux uncertain
by as much as four orders of magnitude. Indeed, detection of WIMP halo
signals may serve to provide information about the DM distribution in
our galaxy.

Anomalies have been reported in the cosmic gamma ray spectrum.  In one
example, the Egret experiment\cite{egret} sees an excess of gamma rays
with $E_\gamma >1$ GeV. Explanations for the Egret GeV anomaly
range from $\tz_1\tz_1\to b\bar{b}\to\gamma$ with 
$m_{\tz_1}\sim 60$ GeV\cite{deboer}, to
mis-calibration of the Egret calorimeter\cite{stecker}. 
The GLAST gamma ray observatory is scheduled for lift-off in 2008 
and should help resolve this issue, as will
the upcoming LHC searches\cite{egret_bbs}.

\subsection{Gravitino dark matter}

In gravity-mediated SUSY breaking models, gravitinos typically have weak
scale masses and, because they only have tiny gravitational couplings,
are usually assumed to be irrelevant for particle physics phenomenology. 
Cosmological considerations, however, lead to the {\it gravitino problem},
wherein overproduction of gravitinos, followed by their late decays
into SM particles, can disrupt the successful predictions of Big Bang
nucleosynthesis. 
The gravitino problem can be overcome by choosing an 
appropriate range for $m_{\tG}$ and a low enough re-heat
temperature for the universe after inflation\cite{gravitinop} as
illustrated in Fig.~\ref{fig:gino}, 
or by hypothesizing that the $\tG$ is in fact the stable LSP, and thus
constitutes the DM\cite{primack}.
\begin{figure}[tbh]
\begin{center}
\includegraphics[width=7cm]{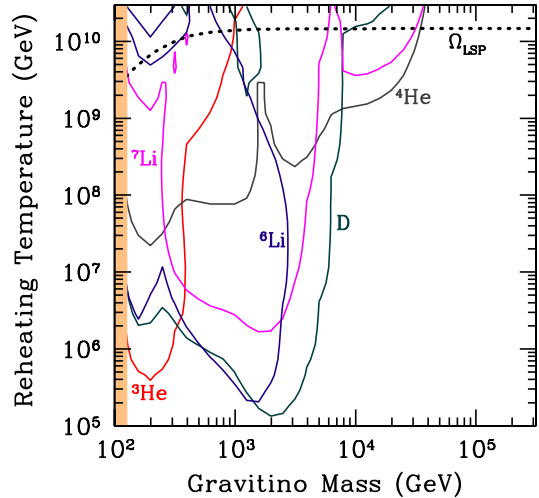}
\end{center}
\vspace{-3mm}
\caption{\small\it An illustration of constraints from Big Bang
nucleosynthesis which require $T_R$ to be below the various curves, for the
HB/FP region of the mSUGRA model with $m_0=2397$ GeV, $m_{1/2}=300$ GeV,
$A_0=0$ and $\tan\beta=30$, from Kohri {\it et al.}\cite{gravitinop} to
which we refer the reader for more details.}
\label{fig:gino} 
\end{figure}

Here, we consider the consequences of a gravitino LSP in SUGRA models. 
If gravitinos are produced in the pre-inflation epoch, then
their number density will be diluted away during inflation. After the
universe inflates, it enters a re-heating period wherein all particles
can be thermally produced. However, the couplings of the gravitino are
so weak that though gravitinos can be produced by the particles that do
partake of thermal equilibrium, gravitinos themselves never attain
thermal equilibrium: indeed their density is so low that gravitino
annihilation processes can be neglected in the calculation of their
relic density.  The thermal production (TP) of gravitinos in the early
universe has been calculated, and including EW contributions, is given
by the approximate expression (valid for $m_{\tG}\ll M_i$\cite{thermal_G}):
\be
\Omega_{\tG}^{TP}h^2\simeq 0.32\left(\frac{10\ GeV}{m_{\tG}}\right)
\left(\frac{m_{1/2}}{1\ {\rm TeV}}\right)^2\left(\frac{T_R}{10^8\
  {\rm GeV}}\right)\; 
\ee
where $T_R$ is the re-heat temperature.

Gravitinos can also be produced by decay of the next-to-lightest SUSY
particle, the NLSP.  In the case of a long-lived neutralino NLSP, the
neutralinos will be produced as usual with a thermal relic abundance in
the early universe. Later, they will each decay as $\tz_1\to \gamma
\tG,\ Z\tG$ or $h\tG$. The total relic abundance is then
\be
\Omega_{\tG}h^2 =\Omega_{\tG}^{TP}h^2+\frac{m_{\tG}}{m_{\tz_1}}\Omega_{\tz_1}h^2.
\ee
The $\tG$ from NLSP decay may constitute warm/hot dark matter depending in the
$\tz_1 -\tG$ mass gap, while the thermally produced $\tG$ 
will be CDM\cite{jlm}.

The lifetime for neutralino  decay to the photon and a gravitino is
given by \cite{fst}, 
\bea
\hspace{-5mm}\tau (\tz_1\to \gamma\tG ) \simeq {{48\pi M_P^2}\over
  m_{\tz_1}^3} A^2 {{r^2}\over{(1-r^2)^3(1+3r^2)}} \nonumber \\
\hspace{1mm}\sim 5.8\times 10^8 \ {\rm s} \left({100 \ {\rm GeV}}\over
{m_{\tz_1}}\right)^3
%  \hspace{-15mm}
{1\over
  A^2} {{r^2}\over{(1-r^2)^3(1+3r^2)}}\;, 
\label{gravdec}
\eea where $A=(v_4^{(1)}\cos\theta_W+v_3^{(1)}\sin\theta_W)^{-1}$, with
$v_{3,4}^{(1)}$ being the wino and bino components of the
$\tz_1$\cite{wss}, $M_P$ is the reduced Planck mass, and
$r=m_{\tG}/m_{\tz_1}$. Similar formulae (with different mixing angle and
$r$-dependence) hold for decays to the gravitino plus a $Z$ or $h$
boson.  We see that -- except when the gravitino is very much lighter
than the neutralino as may be the case in GMSB models with a low SUSY
breaking scale -- the NLSP decays
well after Big Bang nucleosynthesis.  Such decays would inject high
energy gammas and/or hadrons into the cosmic soup post-nucleosynthesis,
which could break up the nuclei, thus conflicting with the successful BBN
predictions of Big Bang cosmology.  For this reason, gravitino LSP 
scenarios usually favor a stau NLSP, since the BBN constraints in this
case are much weaker.

Finally, we remark here upon the interesting interplay of baryogenesis
via leptogenesis with the nature of the LSP and NLSP. For successful
thermal leptogenesis to take place, it is found that the reheat temperature of
the universe must exceed $\sim 10^{10}$ GeV\cite{buchmuller}.  If this
is so, then gravitinos would be produced thermally with a huge
abundance, and then decay late, destroying BBN predictions. For this
reason, some adherents of leptogenesis tend to favor scenarios with a
gravitino LSP, but with a stau NLSP\cite{buchm}.

\subsection{Axion/axino dark matter}

If we adopt the MSSM as the effective theory below $M_{\rm GUT}$, and
then seek to solve the strong $CP$ problem via the Peccei-Quinn solution
\cite{pqww}, we must introduce not only an axion but also a
spin-${1\over 2}$ {\it axino} $\ta$ into the theory.  The axino mass is
found to be in the range of keV-GeV\cite{axmass}, 
but its coupling is suppressed by
the Peccei-Quinn breaking scale $f_a$, which is usually taken to be of
order $10^9-10^{12}$ GeV: thus, the axino interacts more weakly than a
WIMP, but not as weakly as a gravitino.
Both the axion and axino can be compelling choices for DM in the universe\cite{roszk}.

Like the gravitino, the axino will likely not be in thermal equilibrium
in the early universe, but can still be produced thermally via particle
scattering. The thermal production abundance is given
by\cite{roszk,relic_axino} 
\bea \Omega_{\ta}^{TP}h^2 & \simeq & 5.5
g_s^6\log\left(\frac{1.108}{g_s}\right) \left(\frac{10^{11}\ {\rm
GeV}}{f_a/N}\right)^2\nonumber \\ &\times &\left(\frac{m_{\ta}}{100\
{\rm MeV}}\right)\left(\frac{T_R}{10^4\ {\rm GeV}}\right) , 
\eea 
where
$f_a$ is the PQ scale, $N$ is a model-dependent color anomaly factor
that enters only as $f_a/N$, and $g_s$ is the strong coupling at the
reheating 
scale.

Also like the gravitino, the axino can be produced non-thermally by NLSP
decays, where the NLSP abundance is given by the standard relic density
calculation. Thus,
\be
\Omega_{\ta}h^2=
\Omega_{\ta}^{TP}h^2+\frac{m_{\ta}}{m_{NLSP}}\Omega_{\rm NLSP}h^2 .
\ee
In this case, the thermally produced axinos will be CDM for $m_{\ta}\agt
0.1$ MeV\cite{roszk}, while the axinos produced in NLSP decay will constitute
hot/warm DM\cite{jlm}. 
Since the PQ scale is considerably lower than the Planck
scale, the lifetime for decays such as $\tz_1\to \gamma \ta$ are of
order $\sim 0.01-1$ sec-- well before BBN. Thus, the axino DM scenario is
much less constrained than gravitino DM. 

Note also that if axinos are
the CDM of the universe, then models with very large
$\Omega_{\tz_1}h^2\sim 100-1000$ can be readily accommodated, since
there is a huge reduction in relic density upon $\tz_1$ decay to the
axino. This possibility occurs in models with multi-TeV scalars
(and hence a multi-TeV gravitino) and a bino-like $\tz_1$. 
In this case-- with very large $m_{\tG}$-- there is no gravitino
problem as long as the re-heat temperature $T_R\sim 10^6-10^8$ GeV.
This  range of $T_R$  is also what is needed to obtain successful
{\it non-thermal} leptogenesis (involving heavy neutrino $N$
production via inflaton decay)\cite{ntlepto} along with the 
correct abundance of axino dark matter\cite{bs}. 

\subsubsection{Yukawa-unified SUSY with mixed axion/axino dark matter}

The gauge group $SO(10)$ is very highly motivated in that it unifies 
all matter particles of each generation into a spinorial ${\bf 16}$
dimensional representation. Included in the ${\bf 16}$ is a state
containing a SM gauge singlet right-hand neutrino. 
SUSY $SO(10)$ theories may also allow for third generation Yukawa 
coupling unification. In Ref. \cite{yuk,bdr}, it was found via scans over
SUSY parameter space that in fact it is possible to find models with Yukawa
coupling unification, but only for certain choices of GUT scale SSB boundary 
conditions. Yukawa unified solutions can be found if
\bi
\item $m_{16}\sim 3-15$ TeV,
\item $m_{1/2}$ is very small,
\item $\tan\beta \sim 50$,
\item $A_0\sim -2 m_{16}$ with $m_{10}\sim 1.2 m_{16}$,
\item $m_{H_d}>m_{H_u}$ at $M_{GUT}$.
\ei 
In this case, models with an inverted scalar mass hierarchy are found. 
The spectra is characterized by 1. first/second generation scalars in the
3-15 TeV range, 2. third generation scalars, $m_A$ and $\mu$ in the few TeV range and
3. gluinos around 350-500 GeV, with charginos $\sim 100-160$ GeV and 
$\tz_1\sim 50-90$ GeV. 

A problem emerges in that the calculated neutralino relic density is about
$10^2-10^5$ times its measured value. A solution has been invoked that the
axino and axion are instead the DM particles. In this case, neutralinos would still
be produced at large rates in the early universe, but each neutralino would decay
after a fraction of a second (slightly before the onset of BBN) to axino $\ta$ plus
photon. Then the relic axino density would be $(m_{\ta}/m_{\tz_1})\Omega_{\tz_1}h^2$, and the
mass ratio out in front reduces the relic density by a factor of $10^{-2}-10^{-5}$!.
The neutralino is still long lived enough that it will give rise to missing energy at the 
LHC. In fact, since gluinos are so light, we would expect a gluino pair cross section
of order $10^5$ fb, along with decays $\tg\to b\bar{b}\tz_2$, $b\bar{b}\tz_1$ and
$t\bar{b}\tw_1+c.c.$\cite{bkss_lhc}. These new physics reactions should be easily seen via
isolated multi-muon plus jet production in the early stages of LHC running.

The Yukawa-unified SUSY scenario is also very appealing cosmologically.
We would expect the gravitino mass to be of order $m_{16}$: 3-15 TeV. If it is heavier than
about 5 TeV, then the gravitino decays with lifetime around 1 sec: right at the onset
of BBN! This eliminates the BBN gravitino problem as long as re-heat temperature
$T_R\alt 10^9$ GeV. Several scenarios of mixed axion/axino dark matter in Yukawa
unified SUSY have been examined in Refs. \cite{bs,bhkss}. It is found that
one may accommodate either dominant axion or axino DM, but $m_{16}$ at the high end.
Values of $m_{16}\agt 10$ TeV are preferred, as are high values for the PQ breaking scale:
$f_a\agt 10^{12}$ GeV. The values of $T_R$ allowed can accommodate several 
baryogenesis mechanisms, for instance, non-thermal leptogenesis.

\section{Conclusions}

The union of particle physics, astrophysics and cosmology has reached an
unprecedented stage.
Today we are certain that the bulk of the matter in the universe is
non-luminous,  not
made of any of the known particles, but instead made of
one or more {\it new physics} particles that do not appear in the SM.
And though we know just how much of this unknown dark matter there is,
we have no idea {\it what} it is.  Today, many
theoretical speculations seek to answer one of the most pressing
particle physics puzzles, ``What is the origin of EWSB and how can we 
embed this into a unified theory of particle interactions?''
The answer may automatically also point to a resolution of this 75 year old puzzle as to
what the dominant matter component of our universe might be.  Particle
physicists have made many provocative suggestions for the origin of DM,
including supersymmetry and extra spatial dimensions, ideas that will
completely change the scientific paradigm if they prove to be right.

The exciting thing is that many of these speculations will be 
{\it directly tested} by a variety of particle physics experiments along with
astrophysical and cosmological searches. The Large Hadron Collider,
scheduled to commence operation in 2009, will directly study particle
interactions at a scale of 1~TeV where new matter states are anticated
to exist for sound theoretical reasons.  These new states may well be
connected to the DM sector, and so in this way the LHC can make crucial
contributions to not only particle physics, but also to cosmology.
If indeed the LHC can make DM particles or their associated new physics states, 
then a large rate for signal events with jets, leptons and $\eslt$
is expected.

Any discovery at LHC of new particles at the TeV scale will make a
compelling case for the construction of a lepton collider to study the
properties of these particles in detail and to elucidate the underlying
physics. Complementary to the LHC, there are a variety of searches for
signals from relic dark matter particles either locally or dispersed
throughout the galactic halo. The truly unprecedented thing about this
program is that if our ideas connecting DM and the question of EWSB are
correct, measurements of the properties of new particles produced at the
LHC (possibly complemented by measurements at an electron-positron
linear collider) may allow us to independently infer just how much DM
there is in the universe, and quantitatively predict what other searches
for DM should find.

\section{Acknowledgments}

I take this opportunity to thank Tao Han for organizing an exceptional TASI workshop, and
K. T. Mahantappa for his hospitality and patience in organizing many TASI workshops.

\section{References}

%\begin{verbatim}

%\end{verbatim}
\end{document}

%% file: mydefs.tex
\def\CS{{\cal{S}}}

\def\sr2{\sqrt{2}}
\def\CL{{\cal{L}}}
\def\CF{{\cal{F}}}
\def\CM{{\cal{M}}}
\def\CN{{\cal{N}}}
\def\CD{{\cal{D}}}
\def\xhat{\hat{x}}
\def\Shat{\hat{\cal{S}}}
\def\thb{\bar{\theta}}
\def\th{\theta}

\def\dmu{\partial_{\mu}}
\def\gmu{\gamma_{\mu}}

\def\g5{\gamma_5}
\def\ab{\bar{\alpha}}
\def\eslt{\not\!\!{E_T}}

\def\dsl{\not\!\!{\partial}}
\def\to{\rightarrow}

\def\te{\tilde e}

\def\ta{\tilde a}

\def\tu{\tilde u}

\def\tb{\tilde b}
\def\tf{\tilde f}
\def\td{\tilde d}

\def\tG{\tilde G}

\def\psib{\bar\psi}
\def\tst{\tilde t}
\def\ttau{\tilde \tau}

\def\tg{\tilde g}
\def\tnu{\tilde\nu}
\def\tell{\tilde\ell}
\def\tq{\tilde q}
\def\tw{\widetilde W}
\def\tz{\widetilde Z}
\def\alt{\stackrel{<}{\sim}}
\def\agt{\stackrel{>}{\sim}}
\newcommand{\be}{\begin{equation}}
\newcommand{\ee}{\end{equation}}
\newcommand{\bea}{\begin{eqnarray*}}
\newcommand{\eea}{\end{eqnarray*}}